\documentclass[aps,prd,twocolumn,superscriptaddress,showpacs,amsmath,amssymb,amsthm,nofootinbib, preprintnumbers]{revtex4-2}
 \usepackage{amsmath}
\usepackage{graphicx}
\usepackage{epstopdf}
\usepackage{float}
\usepackage{hyperref}
\usepackage{color}
\usepackage[T1]{fontenc}
\usepackage[utf8]{inputenc}
\usepackage[toc,page]{appendix}
\usepackage[usenames,dvipsnames]{xcolor}
\usepackage[normalem]{ulem}
\usepackage{lipsum, babel}

\newcommand{\be}{\begin{equation}}
\newcommand{\ee}{\end{equation}}
\newcommand{\ba}{\begin{eqnarray}}
\newcommand{\ea}{\end{eqnarray}}

\newcommand{\beq}{\begin{equation}}
\newcommand{\eeq}{\end{equation}}
\newcommand{\beqa}{\begin{eqnarray}}
\newcommand{\eeqa}{\end{eqnarray}}

\DeclareMathOperator{\sgn}{sgn}

\begin{document}
%%%%%%%%%%%%%%%%%%%%%%%%%%%%%%%%%%%%%%%%%%%%%%%%%%%%%%%%%%%%%%%%%%%%%%%%%%%%%%%%%%%%%%%%%%

%\title{RegMax electrodynamics: Solutions  and basic properties} 

\title{Solutions and basic properties of regularized Maxwell theory}

\author{Tom{\'a}{\v s} Hale}
\email{tomas.hale@utf.mff.cuni.cz}
\affiliation{Institute of Theoretical Physics, Faculty of Mathematics and Physics,
Charles University, Prague, V Hole{\v s}ovi{\v c}k{\' a}ch 2, 180 00 Prague 8, Czech Republic}

\author{David Kubiz\v n\'ak}
\email{david.kubiznak@matfyz.cuni.cz}
\affiliation{Institute of Theoretical Physics, Faculty of Mathematics and Physics,
Charles University, Prague, V Hole{\v s}ovi{\v c}k{\' a}ch 2, 180 00 Prague 8, Czech Republic}

\author{Otakar Sv{\'i}tek}
\email{ota@matfyz.cz}
\affiliation{Institute of Theoretical Physics, Faculty of Mathematics and Physics,
Charles University, Prague, V Hole{\v s}ovi{\v c}k{\' a}ch 2, 180 00 Prague 8, Czech Republic}

\author{Tayebeh Tahamtan}
\email{tahamtan@utf.mff.cuni.cz}
\affiliation{Institute of Theoretical Physics, Faculty of Mathematics and Physics,
Charles University, Prague, V Hole{\v s}ovi{\v c}k{\' a}ch 2, 180 00 Prague 8, Czech Republic}

%\date{\today}

\date{March 29, 2023}

\begin{abstract}
The regularized Maxwell  theory is a recently discovered theory of non-linear electrodynamics that admits many important gravitating solutions within the Einstein theory. Namely, it 
was originally derived as 
the unique non-linear electrodynamics (that  depends only on the field invariant $F_{\mu\nu}F^{\mu\nu}$) whose radiative solutions can be found in the Robinson--Trautman class. At the same time, it is the only electrodynamics of this type (apart from Maxwell) whose slowly rotating solutions are fully characterized by the electrostatic potential. 
In this paper, after discussing the   basic properties of the regularized Maxwell theory, 
we concentrate on its spherical electric solutions. These not only provide `the simplest' regularization of point electric field and its self-energy, but also feature complex thermodynamic behavior (in both canonical and grandcanonical ensembles) and admit an unprecedented phase diagram with multiple first-order, second-order, and zeroth-order  phase transitions.   
Among other notable solutions, we construct  a novel C-metric describing accelerated AdS black holes in the regularized Maxwell theory. We also 
present a generalization of the regularized Maxwell Lagrangian applicable to magnetic solutions, and find the corresponding spherical, slowly rotating, and weakly NUT charged solutions. 
\end{abstract}

\maketitle

%----------------------------------------------------------------------------------------%
\section{Introduction}
Models of Non-Linear Electrodynamics (NLE) have a long history and reasons 
for considering them evolved throughout the years. The original motivation
came from the efforts to remove divergences associated with point charges and their infinite self-energy.
This led to the discovery of the famous Born--Infeld model \cite{Born:1934gh}, which imposes maximal value for the field strength and yields finite self energy of point charges. (See also \cite{Hoffmann:1937noa} for a ``more regular alternative" to the Born--Infeld theory which yields a point charge with vanishing electric field in the origin.) 
As discovered much later, apart from having unique geometrical and physical properties
\cite{plebanski1970lectures}, 
Born--Infeld-type Lagrangians also arise at low energy regime of string theory \cite{Fradkin:1985qd} and in D-brane physics \cite{Leigh:1989jq}. 
Even more recently certain models of NLE were used to source the so-called regular black holes \cite{Ayon-Beato:1998hmi}, increasing the interest of strong gravity community in NLEs.

To evaluate merits of different NLE models one can employ various criteria -- whether the weak field limit gives rise to Maxwell's theory, potential regularization of a point charge field, conformality, electro-magnetic duality, absence of birefringence etc.  From this point of view the Born--Infeld Lagrangian and the recently derived ModMax theory \cite{Bandos:2020jsw,Kosyakov:2020wxv} that stand out. 
However, when gravity is taken into account, yet another criterion arises. Besides regularizing the black hole geometry {\`a} la \cite{Ayon-Beato:1998hmi}, one may demand that the model be ``consistent with" some important solutions that go {\em beyond spherical} symmetry, such as accelerated black holes, black holes with rotation, or Robinson--Trautman spacetimes.
The NLE model studied in this paper %, which we call the {\em \tcr{Regularized Maxwell (RegMax) theory}},  
is exceptional regarding this latter criterion.

The current model %RegMax model 
was first derived in \cite{Tahamtan:2020lvq}. It admits the Maxwell limit and, as we shall see, provides in some sense the simplest regularization of point charge field and its self-energy. For this reason we call it a {\em Regularized Maxwell (RegMax) theory}. However, the main distinguishing feature of RegMax centers around the fact that it provides important gravitating solutions. 
Namely, 
apart from Maxwell, it is a unique model of NLE (that depends only on the field invariant $F_{\mu\nu}F^{\mu\nu}$) that provides radiative solutions in the Robinson--Trautman class \cite{Tahamtan:2020lvq} (generalizing previous nonradiative results in \cite{Tahamtan:2015bha}). Remarkably, and unlike their Maxwell cousins, such solutions are in addition well-posed \cite{Tahamtan:2020lvq}. 
RegMax theory also provides slowly rotating black holes that can be found  in a form  naturally generalizing the corresponding Maxwell solution \cite{Kubiznak:2022vft}, and, as we shall see shorty, one can also find a very natural generalization of the C-metric -- describing charged accelerated black holes in the RegMax theory.

Our paper is organized as follows. In Sec.~\ref{sec2} we review the basic framework for theories of nonlinear electrodynamics. In Sec.~\ref{sec3} the RegMax theory is introduced and its basic properties are discussed. Sec.~\ref{sec4} is devoted to (electrically charged) spherical solutions and their thermodynamic properties in both canonical and grandcanonical ensembles.
It is shown that such thermodynamics is quite rich, and in particular leads to an  unprecedented grandcanonical phase diagram with multiple phase transitions of various kinds. Other notable electric solutions, including the novel AdS C-metric, are studied in Sec.~\ref{sec5}. Sec.~\ref{sec6} is devoted to a generalization of the RegMax theory that is also applicable to purely magnetic solutions, allowing us to construct magnetically charged black holes, and their slowly rotating and weakly NUT charged cousins.  We summarize our findings in Sec.~\ref{sec7}. App.~\ref{appA} contains formulae for the Maxwell charged and the RegMax charged C-metrics written in the ``standard" C-metric $x-y$ coordinates, drawing the parallel between the two cases.

\section{Theories of nonlinear electrodynamics}
\label{sec2}
A theory of NLE that is 
minimally coupled to Einstein's gravity is  derived from the following action:
\begin{equation}\label{bulkAct}
    I= \frac{1}{16\pi} \int_{M} d^4x \sqrt{-g}\left(R +4{\cal L}-2\Lambda\right)\,,
\end{equation}
where we also included a 
possibility for the cosmological constant $\Lambda$, which we parametrize as 
\be
\Lambda=-\frac{3}{\ell^2}\,, 
\ee
in terms of the AdS radius $\ell$.\footnote{While in this paper we predominantly concentrate on the negative cosmological constant, the de Sitter case can formally be obtained by Wick rotating $\ell$, and the asymptotically flat case by setting $\ell\to \infty$.}
Here, ${\cal L}$ is the electromagnetic Lagrangian, which is taken to be a function of the two electromagnetic invariants 
\be
{\cal S}=\frac{1}{2}F_{\mu\nu}F^{\mu\nu}\,,\quad 
{\cal P}=\frac{1}{2}F_{\mu\nu}(*F)^{\mu\nu}\,,
\ee
where, as always, we have $F_{\mu\nu}=\partial_\mu A_\nu-\partial_\nu A_\mu$, in terms of the vector potential $A_\mu$. 
In order ${\cal L}$ is a true scalar, we have to require that 
\be
{\cal L}={\cal L}({\cal S},{\cal P}^2)\,. 
\ee
Moreover, one might require that the theory of NLE should approach that of Maxwell
\be
{\cal L}_{\mbox{\tiny M}}=-\frac{1}{2}{\cal S}
\ee
in the weak field approximation, 
a condition known as the {\em principle of correspondence}. This condition is satisfied by the RegMax theory studied in this paper, c.f. \eqref{POCRegMax} below.

Introducing the following notation:
\be
{\cal L}_{\cal S}=\frac{\partial {\cal L}}{\partial {\cal S}}\,,\quad  
{\cal L}_{\cal P}=\frac{\partial {\cal L}}{\partial {\cal P}}\,,
\ee
the {\em generalized Maxwell} equations read
\be\label{FE}
d*D=0\,,\quad
dF=0\,, 
\ee
where 
\be\label{Edef}
D_{\mu\nu} = \frac{\partial \mathcal{L}}{\partial F^{\mu\nu}}
=2\Bigl({\cal L_S}F_{\mu\nu}+{\cal L_P}*\!F_{\mu\nu}\Bigr)\, 
\ee
is sometimes referred to as the {\em constitutive relation}.
We also obtain the following 
{\em Einstein equations}: 
\be \label{Hmunu}
G_{\mu\nu}+\Lambda g_{\mu\nu}=8\pi T_{\mu\nu}\,,
\ee
where the generalized electromagnetic energy-momentum tensor reads 
\be\label{Tmunu}
T^{\mu\nu}=-\frac{1}{4\pi}\Bigl(2F^{\mu\sigma}F^{\nu}{}_\sigma {\cal L_S}+{\cal P}{\cal L_P} g^{\mu\nu}-{\cal L}g^{\mu\nu}\Bigr)\,.
\ee

Sometimes, a {\em restricted class} of NLE theories, obtained by considering only the invariant ${\cal S}$: 
\be\label{restrict}
{\cal L}={\cal L}({\cal S})\,, 
\ee
is considered.  This will be the case of the NLE theory studied in this paper. The corresponding equations of motion straightforwardly follow from the above. 

A famous example of a privilaged theory of NLE is the Born--Infeld theory \cite{Born:1934gh}, defined by the following Lagrangian:
\be\label{BI}
{\cal L}_{\mbox{\tiny BI}}=b^2\Bigl(1-\sqrt{1+\frac{\cal S}{b^2}-\frac{{\cal P}^2}{4b^4}}\Bigr)\,. 
\ee
Another recently popular NLE is that of the ModMax theory, discovered in \cite{Bandos:2020jsw, Kosyakov:2020wxv}. This is the most general theory that possesses both, the conformal invariance and the electromagnetic duality.  Its Lagrangian reads 
\be\label{ModMax}
{\cal L}_{\mbox{\tiny ModMax}}=-\frac{1}{2}\Bigl({\cal S}\cosh(\gamma)+\sinh\gamma \sqrt{{\cal S}^2+{\cal P}^2}\Bigr)\,.
\ee

%%%%%%%%%%%%%%%%%%%%%%%%%%%%%%%%%%%%%%
%%%%%%%%%%%%%%%%%%%%%%%%%%%%%%%%%%%%%
\section{RegMax theory: basic properties}
\label{sec3}

In what follows, we focus on yet another type of NLE, defined by the following {\em `RegMax' Lagrangian}:
\ba
{\cal L}&=&-2\alpha^4\,\Bigl(1-3\ln(1-s)+\frac{s^3+3s^2-4s-2}{2(1-s)}\Bigr)\,,\label{Tay}\qquad\\
s&\equiv&\Bigl(-\frac{\mathcal{S}}{\alpha^4}\Bigr)^\frac{1}{4}\in (0,1)\,.\label{s}
\ea
The theory is characterized by a dimensionfull parameter $\alpha>0$, $[\alpha^2]=(\mbox{length})^{-1}$, and reduces to the Maxwell case upon setting 
\be
\alpha\to \infty\,. 
\ee
Namely, we have 
\be\label{POCRegMax}
 {\cal L}=-\frac{{\cal S}}{2}+\frac{4}{5}\frac{(-{\cal S})^{5/4}}{\alpha}+\frac{(-{\cal S})^{3/2}}{\alpha^2}+O(\alpha^{-3})\,.
\ee
On the other hand, the limit $\alpha\to 0$ yields the vacuum case.

Since the RegMax theory contains a dimensionfull parameter, it cannot be, similar to the Born--Infeld theory \eqref{BI}, a {\em conformal} field theory. That this is indeed the case can easily explicitly be shown by calculating the trace of the energy momentum tensor \eqref{Tmunu}, employing that   
\be\label{LsLss}
{\cal L}_{\cal S}=-\frac{1}{2(s-1)^2}\,, \quad 
{\cal L}_{\cal SS}=\frac{1}{4 s^3(1-s)^3\alpha^{7/4}}\,,
\ee
and finding that the trace does not vanish, apart from the Maxwell limiting case, $\alpha\to \infty$.

On the other hand, in contrast to the Born--Infeld case \cite{Gibbons:1995cv},  
the new theory does not enjoy the {\em electromagnetic duality}. Indeed, for this to happen, one would require that the constitutive relation \eqref{Edef} remains invariant under the following transformation \cite{Gibbons:1995cv}:
\be
\delta D_{\mu\nu}=(*F)_{\mu\nu}\,,\quad \delta F_{\mu\nu}=(*D)_{\mu\nu}\,. 
\ee
By employing that $\frac{1}{2}(*F)_{\mu\nu}(*F)^{\mu\nu}=-{\cal S}$, such a duality can easily be shown for the Born--Infeld case or the ModMax case, but it no longer holds for the case of the Lagrangian \eqref{Tay}. It would be interesting to probe, whether it is possible to extend this theory, by appropriately including the invariant ${\cal P}$, so that the electromagnetic duality could be restored.

Theories of NLE propagate two degrees of freedom. However, these need not to propagate along the null cones of the spacetime geometry, nor do they have to propagate with the `same speed'. If the latter happens, we say that the theory suffers from    
{\em birefringence}. This phenomenon is a generic feature of NLE theories.   There are only two exceptions to this rule (see however \cite{Russo:2022qvz}). The Maxwell theory for which the two modes propagate with the speed of light with respect to the gravitational background metric, and the Born--Infeld theory, whose two modes follow the null trajectories of the following effective metric \cite{plebanski1970lectures, Novello:1999pg}:  
\be\label{BI-eff-metric}
g^{\mu\nu}_{\mbox{\tiny eff\, BI}}=
(b^2+{\cal S})g^{\mu\nu} - F^{\mu\alpha}F^{\nu}{}_{\alpha}\,.
\ee
In either case,  no birefringence occurs. The contravariant metric \eqref{BI-eff-metric} leads to causal propagation of light-rays for the Born--Infeld theory.

It can be shown \cite{Novello:1999pg}, that 
for restricted theories ${\cal L}={\cal L}({\cal S})$, 
one mode propagates with respect to the spacetime metric $g_{\mu\nu}$, while the second mode propagates with respect to the following effective metric: 
\be
g^{\mu\nu}_{\mbox{\tiny eff}}=\frac{1}{2}g^{\mu\nu}{\cal L}_{\cal S}+{\cal L}_{\cal SS} F^{\mu\alpha}F^\nu{}_{\alpha}\,. 
\ee
Since we want to check if our NLE theory satisfies causality with respect to the propagation of disturbances (or, equivalently, we consider the nature of characteristic surfaces) we are only interested in the null cones of the effective metric $g^{\mu\nu}_{\mbox{\tiny eff}}$. Thus we can switch to conformally equivalent metric
\be\label{eff-conf-metric}
\hat{g}^{\mu\nu}_{\mbox{\tiny eff}}=g^{\mu\nu}+2\frac{{\cal L}_{\cal SS}}{{\cal L}_{\cal S}} F^{\mu\alpha}F^\nu{}_{\alpha}\,,
\ee
and determine the cone structure. Let us consider arbitrary covector $k_{\mu}$ which is null with respect to the spacetime metric $g^{\mu\nu}k_{\mu}k_{\nu}=0$. Due to antisymmetry of the Maxwell tensor we have $F^{\mu\nu}k_{\mu}k_{\nu}=0$, which means that the vector $l^{\nu}=F^{\mu\nu}k_{\mu}$ is either spacelike or null (and proportional to $k^{\nu}$). The second case leads to identical null cones for both the effective optical and spacetime metric. The first case needs to be investigated further by computing the norm of $k_{\mu}$ with respect to \eqref{eff-conf-metric}, which reads 
\be\label{null-norm}
\hat{g}^{\mu\nu}_{\mbox{\tiny eff}}k_{\mu}k_{\nu}=2\frac{{\cal L}_{\cal SS}}{{\cal L}_{\cal S}} l^{\nu}l_{\nu}\,.
\ee
For the RegMax Lagrangian ${\cal L}_{\cal S}<0$ and ${\cal L}_{\cal SS}>0$,  while $l^{\nu}l_{\nu}>0$ by our assumption. Thus the expression \eqref{null-norm} is negative and therefore the null cone of the spacetime metric is contained within the null cone of the optical metric $g^{\mu\nu}_{\mbox{\tiny eff}}$. However this does not mean that we have superluminal propagation since we need to investigate the corresponding covariant form of the optical metrics which determines the direction of null rays. As noted in \cite{Gibbons:2000xe} the nesting of null cones switches when going from contravariant to covariant forms of respective metrics. This results in {\em causal propagation} of the RegMax modes.

Let us now turn to the gravitating solutions of this theory. As we shall see,  these can be found in a form that is very similar to what happens in the linear Maxwell case.

%%%%%%%%%%%%%%%%%%%%%%%%%%%%%%%%%%%%%%
%%%%%%%%%%%%%%%%%%%%%%%%%%%%%%%%%%%%%%%
\section{Spherical solutions}
\label{sec4}

\subsection{Test charge}

\begin{figure}
\begin{center}
%\begin{figure}[h]
	\includegraphics[scale=0.67]{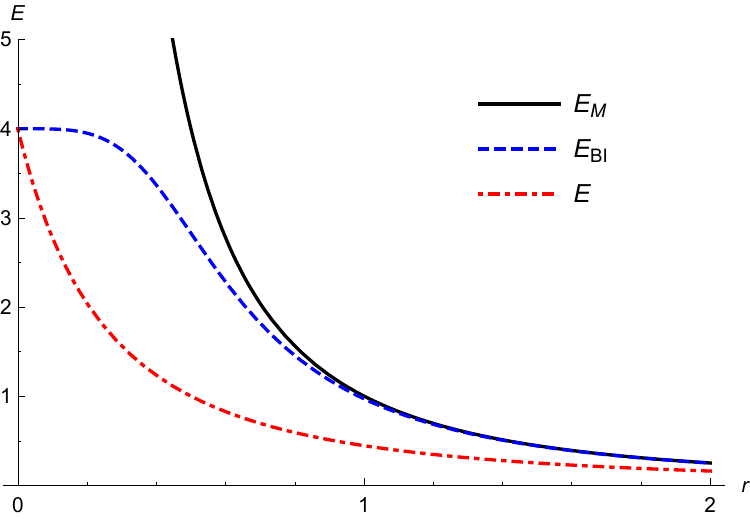}
	\caption{\textbf{Electric field strength.} $E$ of a point charge is displayed for the RegMax theory with $
	\alpha=2$ (red dashdot curve) and compared to the Maxwell (black solid) and Born--Infeld with $b=4$ (dashed blue) cases, setting $Q=1$. } \label{fig1:E} 
\end{center}
\end{figure}

Let us first study the field of a test point-like charge in the RegMax theory \eqref{Tay}, neglecting the backreaction of the electromagnetic field on the geometry. Writing the flat space in spherical coordinates, 
\be
ds^2=-dt^2+{dr^2}+r^2d\Omega^2\,,
\ee
where $d\Omega^2=d\theta^2+\sin^2\!\theta d\varphi^2$, the corresponding solution of RegMax equations  reads
\be\label{phi}
A=\psi_0dt\,,\quad \psi_0=-\frac{\alpha Q}{\alpha r+\sqrt{|Q|}}\,, 
\ee
and is characterized by the following invariants:
\be\label{inv}
{\cal S}=\frac{\alpha^4 Q^2}{(\alpha r+\sqrt{Q})^4}\,,\quad {\cal P}=0\,, 
\ee
where 
\be\label{Qcharge}
Q=\frac{1}{4\pi}\int_{S^2} *D 
\ee
is the (asymptotic) electric charge.

The corresponding field strength
\be\label{E}
F=dA=Edr\wedge dt\,, \quad {E}=\frac{Q\alpha^2}{(\alpha r+\sqrt{|Q|})^2}\,,
\ee
approaches (similar to the Born--Infeld case) a finite value in the origin, $E_0=E(r=0)=\alpha^2 \mbox{ sign}(Q)$.
The electric field \eqref{E} arguably provides the ''simplest regularization'' of a point charge since it gives finite value at the origin while keeping inverse square law profile in the radial coordinate (albeit shifted by a constant).
Moreover, since $E_0$ is finite, the RegMax model leads to finite self-energy, similar to what happens in the Born--Infeld case.

We display the behavior of $E$ in Fig.~\ref{fig1:E}, where it is also compared to the field strength of the Maxwell and Born--Infeld fields, 
\be
E_{\mbox{\tiny M}}=\frac{Q}{r^2}\,,\quad E_{\mbox{\tiny BI}}=\frac{Q}{\sqrt{r^4+Q^2/b^2}}\,, 
\ee
respectively. In this plot we have set $Q=1$ and chosen the value of $\alpha$ and $b$ so that the RegMax and the Born-Infeld field strengths approach the same finite value at the origin.\footnote{Note that since the Born-Infeld parameter $b$ has dimensions of $[b]=(\mbox{length})^{-1}$, whereas the same is true for $\alpha^2$, it is quite natural to compare the values where $\alpha^2\approx b$. Let us also remark that perhaps more physical than comparing the cases with the same finite value at the origin would be to compare the situations with the same integral self-energy of the corresponding charge.}

As is the case for any restricted NLE, the static solution determines the Lagrangian of the theory. In our case, demanding the `simplest regularization' with the `shifted linear profile' \eqref{phi}, uniquely leads to the Lagrangian \eqref{Tay}. While this is not how the theory was originally derived in \cite{Tahamtan:2020lvq} and \cite{Kubiznak:2022vft}, where the consistency of NLE with radiative and slowly rotating solutions was demanded and found to uniquely lead to \eqref{Tay}, we include here the corresponding `derivation' for completeness. Namely,  demanding the regularized solution    
\be
A=-\frac{Q}{r+r_0}dt\,,\quad r_0=\frac{\sqrt{|Q|}}{
\alpha}\,,
\ee
yields the following modified Maxwell equation:
\be
(\nabla \cdot D)_t=\Bigl( 
\frac{2Q{\cal L}_{\cal S}r^2}{(r+\sqrt{|Q|}/\alpha)^2}\Bigr)_{,r}=0\,,
\ee  
that is
\be\label{Lformula}
{\cal L}_{\cal S}=1+\frac{2\sqrt{|Q|}}{\alpha}+\frac{|Q|}{\alpha^2r^2}\,. 
\ee
Moreover, by inverting \eqref{inv}, we get 
\be
r=\frac{\sqrt{|Q|}}{(-{\cal S})^{1/4}}-\frac{\sqrt{|Q|}}{\alpha}\,, 
\ee
which together with \eqref{Lformula}  
leads to \eqref{LsLss}, and upon integration 
to the Lagrangian \eqref{Tay}.

\subsection{RegMax AdS black holes}
The above electrostatic field can easily be upgraded to the full self-gravitating solution of \eqref{bulkAct}.
Such a solution is characterized by a single metric function $f_0$ \cite{Jacobson:2007tj}
and takes the following standard form:
\be\label{SSS}
ds^2=-f_0dt^2+\frac{dr^2}{f_0}+r^2d\Omega^2\,,
\ee
where 
\ba\label{f-SSS}
f_0&=&1-2\alpha^2|Q|+\frac{4\alpha |Q|^{3/2}-6m}{3r}+4r\alpha^3\sqrt{|Q|}\nonumber\\
&&-4\alpha^4r^2\log\Bigl(1+\frac{\sqrt{|Q|}}{r\alpha}\Bigr)+\frac{r^2}{\ell^2}\,,
\ea
while the vector potential $A$ remains `unchanged', given by  \eqref{phi}, with the field strength given by \eqref{E} and field invariants by \eqref{inv}.

\begin{figure}
\begin{center}
%\begin{figure}[h]
	\includegraphics[scale=0.67]{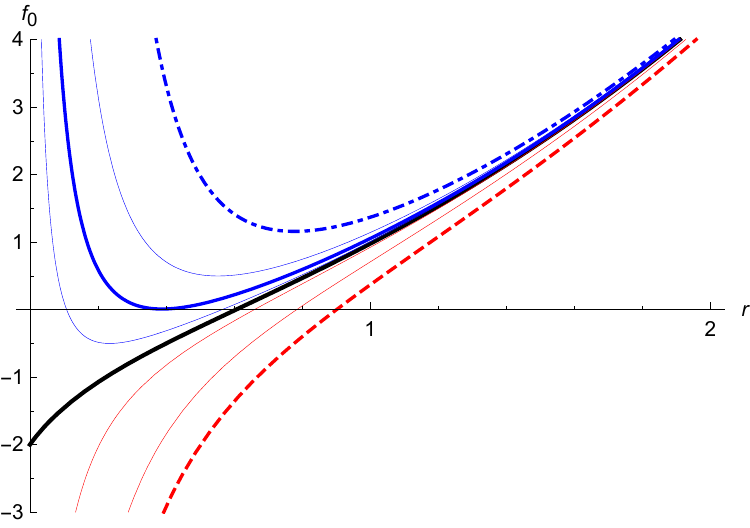}
	\caption{\textbf{Two types of RegMax AdS black holes.} Based on the bahavior of the metric function $f_0$ near the origin, we distinguish two types of RegMax black holes: the RN-type (blue) and the S-type (red). The marginal case $m=M_m$ is highlighted by a thick black line. In particular, for the RN-type ($m<M_m$) we have from top to bottom: naked singularity (top two lines with the top one corresponding to $m\to0$), extremal black hole with one horizon (solid blue), and the case with two horizons (lower thin blue). The figure is displayed for $Q=1$, $\ell=1$, and $\alpha=1$.   
}\label{fig2}
\end{center}
\end{figure}

\begin{figure}
\begin{center}
%\begin{figure}[h]
	\includegraphics[scale=0.67]{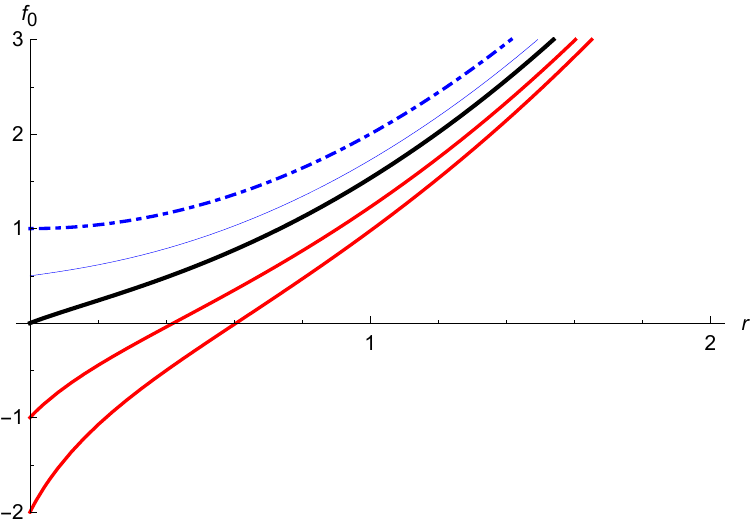}
	\caption{\textbf{Marginal case.} The sequence of marginal metric functions $f_0$ is displayed for various values of $\alpha$: $\alpha=\sqrt{\frac{3}{2}}, 1, \frac{1}{\sqrt{2}}, \frac{1}{2},$ and $\alpha\to0$ %\tcc{\bf just solutions to $(-2,-1,0,1/2,1)=1-2\alpha^2$; for $\alpha\rightarrow0$, $f_0$ is defined only in the limit} 
(from bottom to top) and fixed $Q=1=\ell$. As $\alpha$ decreases, the marginal curve moves up, shifting the RN-branch to more positive values. For $\alpha<\alpha_c=1/\sqrt{2}$ (blue curves) the RN-branch is necessarily always positive and describes a naked singularity; in this case any black hole solution is necessarily described by the S-branch.}\label{fig3}
\end{center}
\end{figure}

The solution possesses a singularity at $r=0$, as can for example be seen by the expansion of the Ricci scalar:
\be
R=-\frac{4|Q|\alpha^2}{r^2}+O(1/r)\,,
\ee
and the Kretschmann scalar:
\ba
{\cal K}&=&R_{\alpha\beta\gamma\delta}R^{\alpha\beta\gamma\delta}\nonumber\\
&=&\frac{16\,\left( 2\alpha |Q|^{3/2}-3m \right)^2}{3r^6}+O(1/r^5)\,.
\ea
Obviously, the singularity prevails irrespective of the choice of the mass parameter $m$, but the divergence of the Kretschmann scalar can be made `milder', compared to the Schwarzschild case with ${\cal K}\sim \frac{1}{r^6}$, by selecting specific values of $m$ and $Q$.

Depending on the choice of parameters, the solution describes a black hole with one or two horizons, or a naked singularity. More concretely, following the discussion for the Born--Infeld case in \cite{Gunasekaran:2012dq}, we expand the metric function around the origin, to obtain 
\be
f=\frac{2(M_m-m)}{r}+1-2|Q|\alpha^2+4\alpha^3\sqrt{|Q|}r+O(r^2)\,, 
\ee
where 
\be
M_m\equiv \frac{2\alpha |Q|^{3/2}}{3}\, 
\ee
is the `marginal mass'. For $m>M_m$ we have a `Schwarzschild-like' (S-type) black hole characterized by a single horizon. On the other hand, when $m<M_m$, the behavior is more `Reissner--Nordstrom-like' (RN-type) and we can have two, one extremal, or no horizons, see Fig.~\ref{fig2}. For the marginal case, $m=M_m$, the metric function approaches a finite value in the origin, 
$f_0(r=0)=1-2|Q|\alpha^2$. When this is positive, that is for 
\be
|Q|\alpha^2<|Q|\alpha_c^2=\frac{1}{2}\,, 
\ee
the Reissner--Nordstrom-like solution is a naked singularity, and only the Schwarzschild-like branch describes a black hole, see Fig.~\ref{fig3}. As we shall see, $\alpha_c$ plays an important role for the thermodynamic behavior of the solution.

In the case when we have a black hole, its horizon is located at the 
largest root $r_+$ of 
$f_0(r_+)=0$. It is a Killing horizon generated by the following Killing field:  
\be
\xi=\partial_t \,.
\ee
Because of the presence of the logarithmic term in \eqref{f-SSS}, the position of $r_+$ has to be determined numerically.

\subsection{Thermodynamics}

Let us next turn to the thermodynamic properties of the obtained solution. The temperature and entropy are given by the standard formulae and read 
\ba
T&=&\frac{f'(r_+)}{4\pi}\nonumber\\
&=&\frac{\alpha r_+(6|Q|\alpha^2+1)-2|Q|^{3/2}\alpha^2+\sqrt{|Q|}(1+12\alpha^4r_+^2)}{4\pi r_+(\alpha r_++\sqrt{|Q|})}\nonumber\\
&&-\frac{3r_+\alpha^4}{\pi}\log\bigl(1+\frac{\sqrt{|Q|}}{r_+\alpha}\bigr)+\frac{3r_+}{4\pi \ell^2}\,,\label{Tform}\\
S&=&\frac{\mbox{Area}}{4}=\pi r_+^2\,.
\ea
The asymptotic electric charge, \eqref{Qcharge}, is given by $Q$, and the electrostatic potential $\phi$ is identified with 
\be\label{phiTD}
\phi=-\xi\cdot A\Bigr|_{r=r_+}=\frac{\alpha Q}{\alpha r_++\sqrt{|Q|}}\,. 
\ee
The thermodynamic mass can be, for example, calculated by the conformal method \cite{Ashtekar:1999jx}, and reads
\be
 M=m\,.
\ee
Finally, since the solution is asymptotically AdS, we can consider the corresponding pressure-volume term \cite{Kastor:2009wy, Kubiznak:2016qmn}, 
\be\label{Vform}
P=-\frac{\Lambda}{8\pi}=\frac{3}{8\pi \ell^2}\,,\quad  V=\Bigl(\frac{\partial M}{\partial P}\Bigr)_{S,Q,
\alpha}=\frac{4}{3}\pi r_+^3\,,
\ee
together with the ``{\em $\alpha$-polarization potential}" \cite{Gunasekaran:2012dq}
\ba
\mu_\alpha&=&\Bigl(\frac{\partial M}{\partial \alpha}\Bigr)_{S,Q,P}\nonumber\\
&=&-\frac{2}{3}\frac{2|Q|^{3/2}\alpha r_+-Q^2-12\alpha^3r_+^3\sqrt{|Q|}-6|Q|\alpha^2r_+^2}{r_+\alpha+\sqrt{|Q|}}\nonumber\\
&&-8\alpha^3 r_+^3\log\Bigl(1+\frac{\sqrt{|Q|}}{r_+\alpha}\Bigr)\,. 
\ea
With these in hand, it is now easy to verify that the extended first law:
\be
\delta M=T\delta S+\phi \delta Q+V\delta P+\mu_\alpha \delta \alpha\, 
\ee
is satisfied. This is accompanied by the Smarr relation, which correspondingly  includes the extra $\alpha \mu_\alpha$ and $PV$ terms:
\be
M=2TS+\phi Q-2VP-\frac{1}{2}\mu_\alpha \alpha\,, 
\ee
reflecting the dimensionality of the corresponding thermodynamic quantities, e.g. \cite{Kastor:2009wy}.

\subsection{Canonical ensemble}

It is well know that in the canonical (fixed $Q$) ensemble charged-AdS black holes in the Maxwell theory feature a first-order small black hole/large black hole phase transition {\`a} la Van der Waals  that terminates at a critical point, characterized by the standard critical exponents \cite{Chamblin:1999tk, Kubiznak:2012wp}. For the Born--Infeld case, the situation is even more interesting \cite{Gunasekaran:2012dq} (see also \cite{Ali:2023wkq}). Namely, while the Van der Waals-like phase transition still exists for large enough $b$, for sufficiently small $b<b_1=1/(\sqrt{8}|Q|)$, the Schwarzschild-like behavior prevails and there is no criticality. At the same time there exists an interesting intermediate region, $b\in (b_1, b_2)$, where $b_2=1/(2|Q|)$, for which the phase behavior features `multicomponent behavior' and associated with it reentrant phase transitions \cite{Gunasekaran:2012dq, Altamirano:2013ane}. As we shall see in this section, such an intermediate region is absent for the black holes in the RegMax theory.

To uncover the thermodynamic behavior of the RegMax solutions, we need to study the  (canonical ensemble) free energy  
\be
F=M-TS=F(T,Q,P,\alpha)\,, 
\ee
whose behavior crucially depends on the value of parameter $\alpha$. In what follows we parametrically plot $F-T$ diagrams for fixed $Q$, and various $\alpha$'s and $P$'s, using $r_+$ as a parameter.

\begin{figure}
\begin{center}
%\begin{figure}[h]
	\includegraphics[scale=0.67]{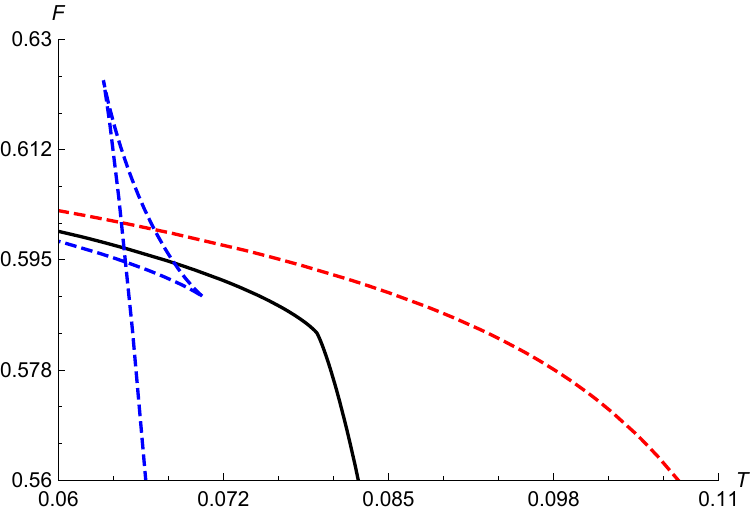}
	\caption{\textbf{$F-T$ diagram: $\alpha=1>\alpha_c$}. The diagram is displayed for various pressures $P$ and fixed $Q=1$. For $P<P_c$ (blue curve) we observe the swallowtail behavior characteristic of the first order phase transition. At $P=P_c$ (solid black) the swallowtail degenerates to a single critical point, giving rise to a second order phase transitiont. Above $P_c$ (red curve), the free energy is smooth and single valued.
}\label{fig4} 
\end{center}
\end{figure}

\begin{figure}
\begin{center}
%\begin{figure}[h]
	\includegraphics[scale=0.67]{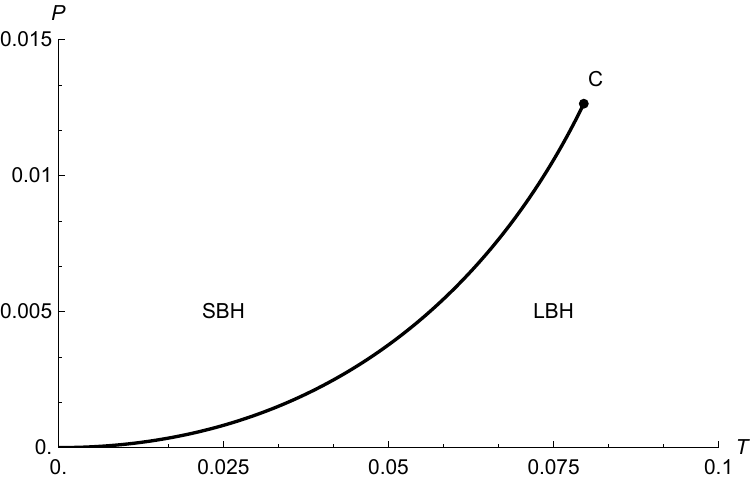}
	\caption{\textbf{$P-T$ phase diagram: $\alpha>\alpha_c$.} For $\alpha>\alpha_c$ the system admits a first order phase transition between small and large black hole phases, reminiscent of the Van der Waals fluid.   The corresponding coexistence line is displayed by thick black curve, which terminates at a critical point denoted by $C$. The phase to the left describes small black holes (SBH), whereas to the right we have the large black hole (LBH) phase. Above $P_c$ one can no longer distinguish the two phases. The figure is plotted for $Q=1$ and $\alpha=1$. }\label{fig5} 
\end{center}
\end{figure}

For  $\alpha>\alpha_c$, we observe the `standard' behavior known from the charged-AdS black holes in Maxwell's theory, see Fig.~\ref{fig4}. Namely, for fixed $Q$ and $\alpha$, there exists a critical pressure $P_c$ below which
the free energy demonstrates the characteristic swallow tail behavior, connected with the {\em small black hole (SBH)/large black hole (LBH)} first order phase transition. At $P=P_c$ this swallow tail degenerates to a single critical point, located at $(P_c, V_c, T_c)$, at which the first order phase transition terminates and becomes of the second order. Above $P_c$, the free energy becomes smooth and single valued, indicating the presence of a single phase. 
The expression for $(P_c, V_c, T_c)$ can be calculated from the standard relations
\be\label{CritPoint}
\frac{\partial P}{\partial V}=0=\frac{\partial^2 P}{\partial V^2}\,, 
\ee
where $P=P(V,T, Q, \alpha)$ is obtained by  rewriting \eqref{Tform} and using \eqref{Vform}. 
However, for a given $\alpha$ and $Q$ \eqref{CritPoint} leads to a fifth order polynomial, and is better solved numerically. 
We display the resulting (numerically constructed) $P-T$ phase diagram in Fig.~\ref{fig5}. It shows the coexistence line between SBH and LBH phases that terminates at a critical point denoted by $C$. We expect that, similar to the Maxwell case \cite{Kubiznak:2012wp, Gunasekaran:2012dq}, $C$ is  characterized by the standard mean field theory critical exponents.

\begin{figure}
\begin{center} 
%\begin{figure}[h]
	\includegraphics[scale=0.67]{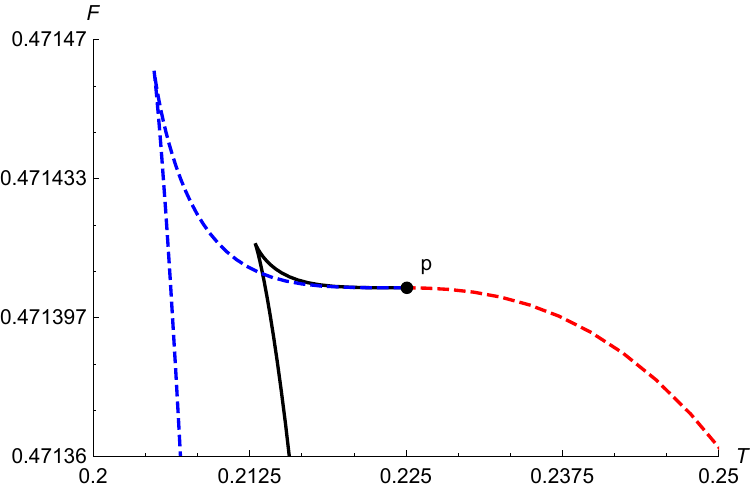}
	\caption{\textbf{$F-T$ diagram: $\alpha=\alpha_c$}, The diagram is displayed for
$\ell=0.4$ (red dashed), $\ell=0.7$ (black), and $\ell=0.8$ (blue dashed) and $Q=1$.	
In this marginal case we observe a special point $p$ at finite $T$ and $F$ (independent of $P$) from where the curves of free energy emerge. While the cusp remains present, the swallow tail is disrupted by the existence of $p$. Consequently the phase diagram features only one (large) black hole phase and a no black hole region for small enough temeperatures. }\label{fig6} 
\end{center}
\end{figure}

\begin{figure}
\begin{center}  
%\begin{figure}[h]
	%\includegraphics[scale=0.5]{NOBH_LBH}
	\includegraphics[scale=0.46]{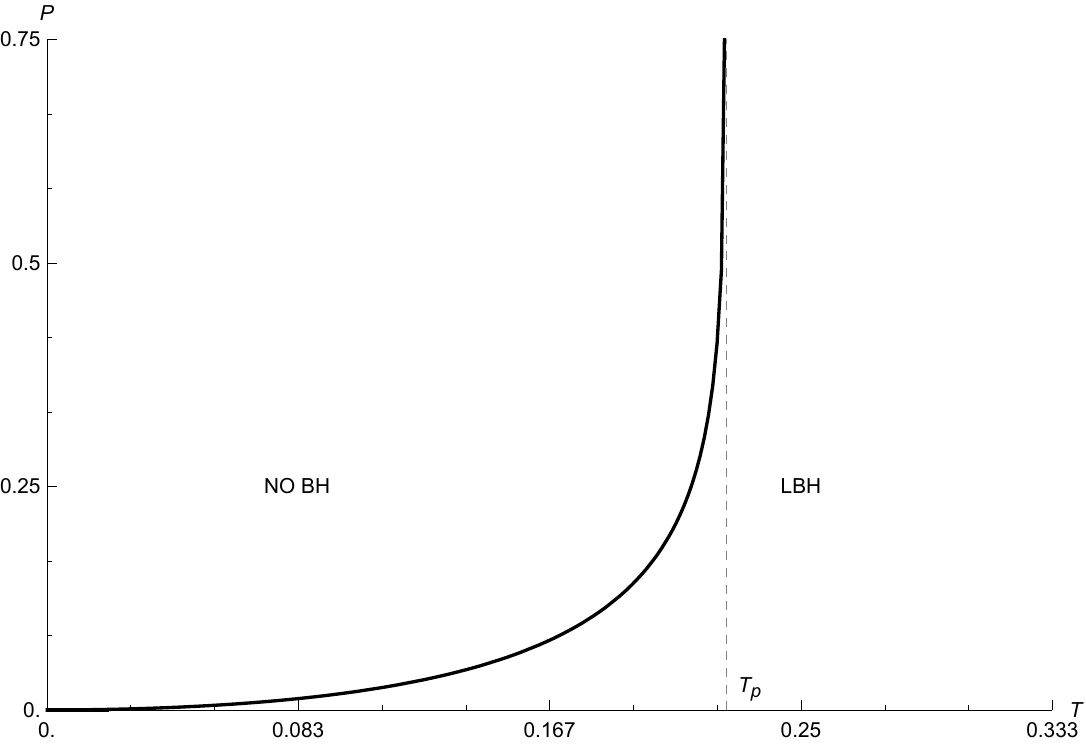}
	\caption{\textbf{$P-T$ phase diagram: $\alpha=\alpha_c$.} In this marginal case, the only black hole phase is that of large black holes (LBH). We also observe a forbidden no black hole region (NO BH). The separation line between the two regions asymptotes to $T_p$, indicated by thin grey line. For $\alpha<\alpha_c$ the phase diagram would look qualitatively similar, but the separation line would extend to arbitrarily high temperatures. } 
\label{fig7} 
\end{center}
\end{figure}

For $\alpha=\alpha_c$, the behavior of the free energy suddenly changes, as displayed in Fig.~\ref{fig6}. Namely, we observe point $p$  from where the free energy emerges at $r_+=0$. The existence of this point `destroys' a possibility for a swallow tail and we no longer have a first order phase transition between small and large black holes. The position of $p$ is independent of pressure, and is easily determined by expanding the corresponding ($\alpha=\alpha_c$) free energy and temperature for small $r_+$:
\ba\label{Tppoint}
T&=&T_p+O(r_+)\,,\quad T_p=\frac{1}{\sqrt{2}\pi |Q|}\,,\nonumber\\ 
F&=&\frac{Q\sqrt{2}}{3}+O(r_+^3)\,. 
\ea 
As the black hole radius increases from $r_+=0$, the free energy curve `heads left' until it meets a cusp, from where it continues to lower free energies and higher temperatures (lower branch).   
As displayed in the figure, as pressure decreases, the cusp occurs at smaller and smaller temperature,   below which black holes no longer can exist -- indicating an onset of a forbidden `{\em no black hole (NO BH)} region'.\footnote{Since we study canonical (fixed $Q$) ensemble, there is no radiation phase with $F\approx 0$ (we cannot have charged radiation) and no Hawking--Page-like phase transition occurs.}  
For $\alpha=\alpha_c$ the systems thus has only one phase of large black holes (LBH) (corresponding to the lower branch of the free energy) and a no black hole region for small enough temperatures, as displayed in 
Fig.~\ref{fig7}. Note also, that the line of separation between NO BH and LBH regions asymptotes to a finite temperature, given by the temperature $T_p$, \eqref{Tppoint}, of the endpoint $p$.

\begin{figure}
\begin{center}
%\begin{figure}[h]
	\includegraphics[scale=0.67]{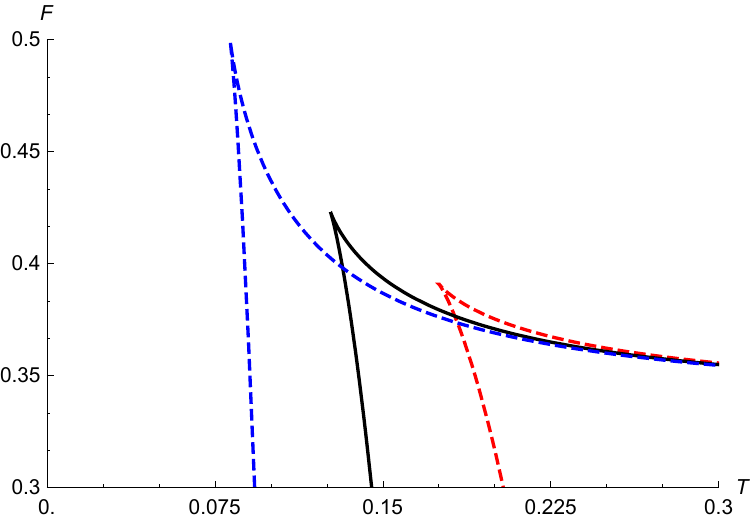}
	\caption{\textbf{$F-T$ diagram: $\alpha<\alpha_c$.} The diagram is displayed for $\alpha=\frac{1}{2}$, $Q=1$, and $\ell=0.4$ (red dashed), $\ell=1$ (black) and $\ell=2.2$ (blue  dashed). The point $p$ no longer exists, and we observe two branches of black holes separated by a cusp that gives rise to a no black hole region. The lower branch corresponds to large black holes and is thermodynamically preferred. }\label{fig8} 
\end{center}
\end{figure}

Finally, for $\alpha<\alpha_c$, we observe $F-T$  behavior shown in Fig.~\ref{fig8}. The point $p$ no longer exists and small black holes (in the upper branch) have arbitrarily large temperature (and positive free energy). We still have a cusp 
at finite temperature from where the (lower) branch of large thermodynamically favoured black holes emerges. This results in a phase diagram that is qualitatively very similar to that in Fig.~\ref{fig7} (NO BH region exists for small enough temperatures whereas LBHs are thermodynamically favoured for large temperatures). The only 
significant difference is that the separation line between NO BH and LBH regions no longer asymptotes to $T_p$ but rather extends to arbitrarily large temperatures.

\begin{figure}
\begin{center}
%\begin{figure}[h]
	\includegraphics[scale=0.67]{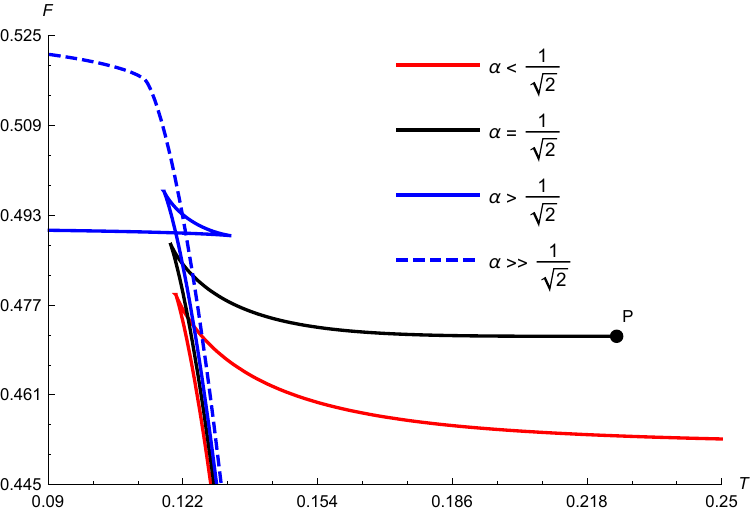}
	\caption{\textbf{$F-T$ diagram: the effect of $\alpha$.} The $F-T$ behaviour is displayed as a function of $\alpha$ for fixed $Q=1$ and fixed pressure $\ell=2$. Contrary to the Born--Infeld case, there is no intermediate region of $\alpha$'s for which the multicomponent behavior exists. Rather, a sharp transition between the Van der Waals-like behavior and the Schwarzschil-like  behavior accurs at $\alpha=\alpha_c$.}\label{fig9}
\end{center} 
\end{figure}

Having described all cases, we see that the 
situation is very different from the Born--Infeld case \cite{Gunasekaran:2012dq}, as we no longer observe an intermediate 
range for $\alpha$, for which the system would feature multicomponent behavior and reentrant phase transitions. Instead, when $\alpha>\alpha_c$ we observe the Van der Waals-like behavior and when $\alpha<\alpha_c$ the behavior is Schwarzschild-like, see Fig.~\ref{fig9} for the display of the free energy at fixed $Q$ and $P$.
The absence of reentrant phase transitions can mathematically be linked to the absence of physical critical points below $\alpha_c$, whereas such critical points do exist below $b_2$ in the Born--Infeld case. 
It remains an interesting open question which type of behavior, whether the one with intermediate region of reentrant phase transitions, or the one without it, is more generic for black holes in nonlinear electrodynamics.

\subsection{Grandcanonical ensemble}
The thermodynamic behavior of RegMax black holes  is also interesting in the grandcanonical (fixed $\phi$) ensemble. In this case, we are interested in 
the grandcannonical free energy:
\be
W=M-TS-\phi Q=W(T,\phi, P, \alpha)\,. 
\ee
The key difference from the canonical ensemble is that we can now have a new phase of (neutral and fixed $\phi$) {\em thermal radiation}, characterized by 
\be 
W\approx 0\,.
\ee
As we shall see, this will result in the occurance of the first-order phase transitions {\`a} la Hawking--Page \cite{Hawking:1982dh}, and the zeroth-order phase transitions \cite{Gunasekaran:2012dq, Altamirano:2013ane} between radiation and black hole phases. 
Similar to what happens in the canonical ensemble, 
we will also observe small to large black hole phase transitions (see  \cite{Hendi:2012um, Zou:2014mha, Sherkatghanad:2014hda, Liang:2019dni, Zhou:2020vzf} for similar  type of grandcanonical behavior previously observed in other non-linear settings). When all such  phenomena are considered together, we arrive at a rather complicated-looking phase diagram \ref{fig16}, clearly demonstrating the complexity of 
thermodynamics of RegMax black holes.

To start with, in order to express $W$ and $T$ in terms of $\phi$ rather than $Q$, we need to invert the relation \eqref{phiTD}. For $\alpha$ positive, this yields a single solution: 
\begin{align}
Q=\frac{\phi}{2\alpha^2}\left(2\alpha^2r_++|\phi|+\sqrt{\phi^2+4|\phi|\alpha^2r_+}\right)\,. 
\end{align}
In what follows we concentrate (without loss of generality) on positive $\phi$. 
Plugging this in formula for the temperature $T$ and expanding for small $r_+$, we find that there exists a critical value for the potential $\phi$:
\be
\phi_c=\frac{1}{\sqrt{2}}\,, 
\ee
at which the temperature `snaps' from plus to minus infinity. This is very similar to what happens for $\alpha_c$ in the canonical ensemble and, as we shall see, has significant implications for the thermodynamic behavior.
Let us first study the corresponding $P-T$ phase diagrams.

\begin{figure}
\begin{center}
%\begin{figure}[h]
	\includegraphics[scale=0.67]{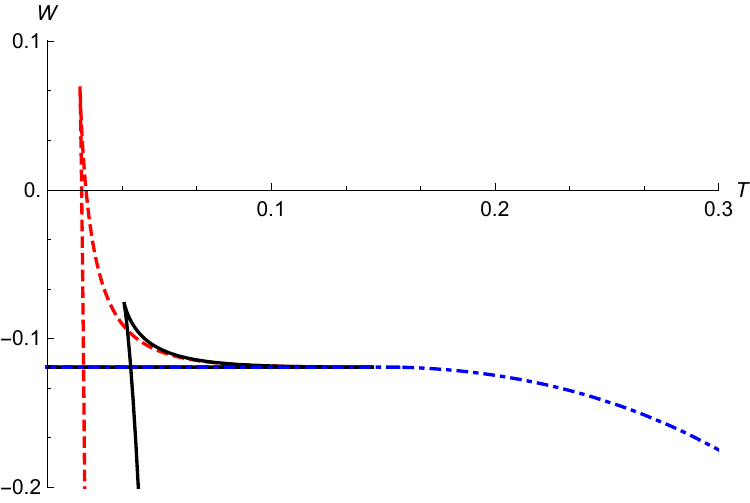}
	\caption{\textbf{$W-T$ diagram: $\phi>\phi_c$.} The grandcanonical free energy $W$ is plotted against the temperature $T$ for constant $\phi=0.71>\phi_c$ and $\alpha=1$ at $P=0.0008$ (red dashed swallowtail), $P=0.005$ (solid black swallowtail), and $P=0.25$ (a smooth decreasing region dispalyed by blue dot-dashed curve). The thermodynamically preferred black holes have negative $W$ and shield the radiation phase at $W\approx 0$.
 }\label{fig10} 
\end{center}
\end{figure}

\begin{figure}
\begin{center}
%\begin{figure}[h]
	\includegraphics[scale=0.67]{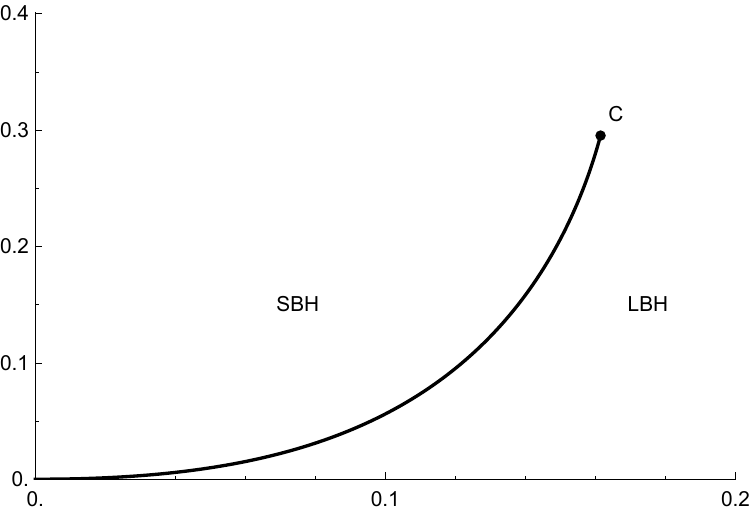}
	\caption{\textbf{$P-T$ diagram: $\phi>\phi_c$.} This phase diagram (displayed for $\phi=0.71$ and $\alpha=1$) features a standard small to large black hole phase transition {\`a} la Van der Waals. Its existence in the grandcanonical ensemble is a direct consequence of the non-linearity of the electromagnetic field.}\label{fig11} 
\end{center}
\end{figure}

For $\phi>\phi_c$, fixed $\alpha$ and various pressures, we observe Van der Waals-like behavior of the $W-T$ diagram, see Fig.~\ref{fig10}. Namely, there exists a critical pressure $P_c$ at which the critical point develops, while we observe swallowtail behavior for $P<P_c$, and smooth free energy above $P_c$. Importantly, thermodynamically preferred black holes have negative $W$  and thence `shield' the radiation phase at $W\approx 0$, which is in this case thermodynamically unfavoured. 
The corresponding phase diagram, displayed in Fig.~\ref{fig11}, thus features a first order small to large black hole phase transition, and it is very similar to what happens in the canonical ensemble, c.f. Fig.~\ref{fig5}. Let us stress, that the very existence of such a phase transition in the grandcanonical ensemble is a direct consequence of non-linearity of the electromagnetic field and it is absent for black holes in the Maxwell theory. It has, however, been observed for black holes in other theories of non-linear electrodynamics and in higher curvature gravities, e.g. \cite{Hendi:2012um, Zou:2014mha, Sherkatghanad:2014hda, Liang:2019dni, Zhou:2020vzf}.

\begin{figure}
\begin{center}
%\begin{figure}[h]
	\includegraphics[scale=0.67]{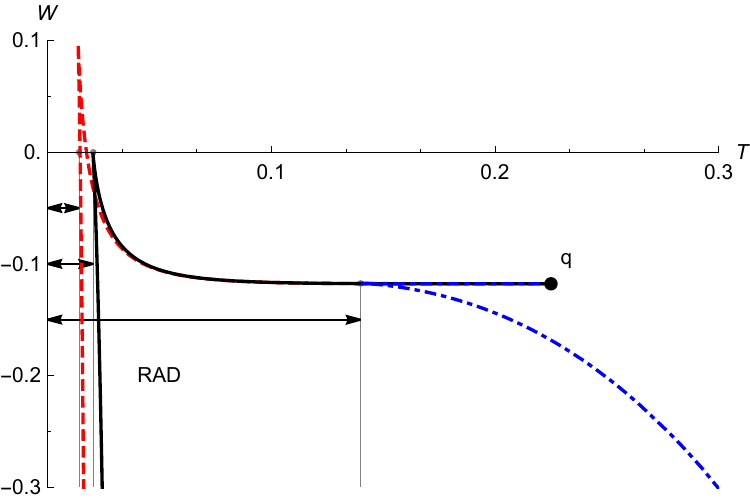}
	\caption{\textbf{Marginal $W-T$ diagram: $\phi=\phi_c$.} The grandcanonical free energy $W$ is plotted against the temperature $T$ for fixed $\phi=\frac{1}{\sqrt{2}}$ and $\alpha=1$, and  $P=0.0007$ (red dashed), $P=0.0016$ (black solid) and $P=0.153$ (blue dash-dot). All these lines terminate at the end point $q$ as $r_+\to 0$.  The area between $T=0$ and each one of the vertical thin grey lines, as indicated by the arrows, is characterized by a radiation phase with $W\approx0$.
	Depending on the position of the cusp, there is either first-order, or zeroth-order phase transition between the radiation phase and the large black hole phase (lower branch). Namely,  for $P\lessapprox0.0016$ (red and black lines) there exists a first order phase transition, while for larger $P$ this becomes a zeroth order phase transition (blue line).	
}\label{fig12} 
\end{center}
\end{figure}

\begin{figure}
\begin{center}
%\begin{figure}[h]
	\includegraphics[scale=0.72]{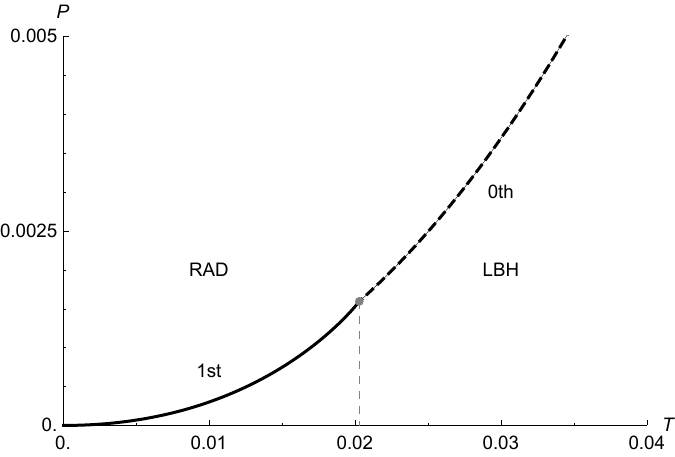}
	\caption{\textbf{$P-T$ diagram: $\phi=\phi_c$.}  For small enough pressures, the phase diagram features a radiation/large black hole first-order phase transition {\`a} la Hawking Page. The corresponding coexistence line terminates at a point for which $W_{\mbox{\tiny cusp}}=0$ from where it `continues' as a coexistence line of the zeroth-order phase transition between the radiation and large black hole phases. This curve eventually asymptotes to the temperature of the endpoint $q$, $T_q$ (not displayed in the figure). For $\phi<\phi_c$ the diagram would be qualitatively similar, except the zeroth-order coexistence line would no longer asymptote to $T_q$ and rather it would evolve to arbitrarily high temperatures. }\label{fig13} 
\end{center}
\end{figure}

The $W-T$ diagram for the marginal case $\phi=\phi_c$ is displayed in Fig.~\ref{fig12}. Similar to what happens at point $p$ in the canonical ensemble, we now observe the `end point' $q$ where all free energy curves terminate as $r_+\to 0$. 
This limit can be taken explicitly and results in
\ba\label{Tqpoint}
	T&=&T_q+O(r_+)\,,\quad T_q=\frac{\alpha^2}{\sqrt{2} \pi }\,,\nonumber\\ 
	W&=&-\frac{1}{6 \sqrt{2} \alpha^2}+O(r_+^3)\,.
\ea
Obviously, the position of $q$ is independent of pressure and only depends on the value of $\alpha$. 
As with the point $p$, the very existence of $q$ `destroys' the swallowtail and only a cusp, located at $(T_{\mbox{\tiny cusp}}, W_{\mbox{\tiny cusp}})$,  remains present. However, since we now also have the radiation phase, at $W\approx 0$, the situation is different from that of the canonical ensemble, and, depending on the position of this cusp, several cases may happen.

Namely, when $W_{\mbox{\tiny cusp}}<0$, which happens for a given $\alpha$ and large enough pressure, $P>P_0(\alpha)$, see blue dash-dot curve in Fig.~\ref{fig12}, there will be a {\em zeroth-order} phase transition at $T=T_{\mbox{\tiny cusp}}$ between the radiation phase, present for $T<T_{\mbox{\tiny cusp}}$ and the large black hole phase (lower branch of $W$) for $T>T_{\mbox{\tiny cusp}}$. As pressure decreases, the cusp `moves up', and at $P=P_0$ it is characterized by $W_{\mbox{\tiny cusp}}=0$, see solid black curve in Fig.~\ref{fig12}. At this point, at $T=T_{\mbox{\tiny cusp}}$, the zeroth-order phase transition between the radiation and large black hole phases terminates and becomes of the first order. As we lower the pressure even further, $W_{\mbox{\tiny cusp}}$ moves to positive values and we observe an {\em analogue of the Hawking--Page} first order phase transition 
at $T=T_{\mbox{\tiny HP}}$ at which the lower branch of the free energy crosses $W=0$. That is, for $T<T_{\mbox{\tiny HP}}$ we have a radiation phase, which at $T=T_{\mbox{\tiny HP}}$ `condenses' to a thermodynamically preferred large black hole phase present for $T>T_{\mbox{\tiny HP}}$. The corresponding $P-T$ phase diagram is summarized in Fig.~\ref{fig13}. Similar to what happens with point $p$, note that, due to the existence of the endpoint $q$, the zeroth-order phase transition coexistence line would eventually asymptote to finite temperature $T_q$, \eqref{Tqpoint}, of the endpoint $q$ (not displayed in the figure).

\begin{figure}
\begin{center}
%\begin{figure}[h]
	\includegraphics[scale=0.67]{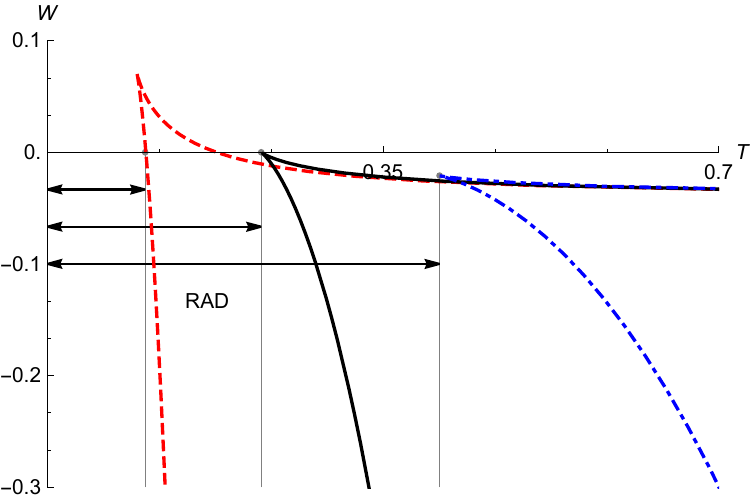}
	\caption{\textbf{$W-T$ diagram: $\phi<\phi_c$.} Depending on the position of the cusp of the grandcanonical free energy $W$, there is either first-order, or zeroth-order phase transition between the radiation phase (characterized by $W=0$) and the large black hole phase (lower branch). The figure is displayed for 
$\phi=0.5$ and $\alpha=1$, and for $P=0.02$ (red dashed), $P=0.12$ (black solid) and $P=0.42$ (blue dot-dashed). The area between $T=0$ and each of the vertical thin grey lines, as indicated by the arrows, is characterized by a radiation phase. We see that for $P\lessapprox0.12$ (red and black lines) there exists a first order phase transition, while for larger $P$ this becomes a zeroth order phase transition (blue line).}\label{fig14} 
\end{center}
\end{figure}

Finally, for $\phi<\phi_c$, we observe the free energy displayed in Fig.~\ref{fig14}. Similar to the previous case, the thermodynamic behavior depends on the position of the cusp, and results in radiation/large black hole zeroth-order phase transition for large enough pressures, and Hawking--Page-like phase transition for small pressures. The corresponding phase diagram is 
qualitatively similar to Fig.~\ref{fig13}, with only only difference that the zeroth-order coexistence line no longer asymptotes to $T_q$ but rather evolves to arbitrarily high temperatures.

\begin{figure}
\begin{center}
%\begin{figure}[h]
	\includegraphics[scale=0.7]
{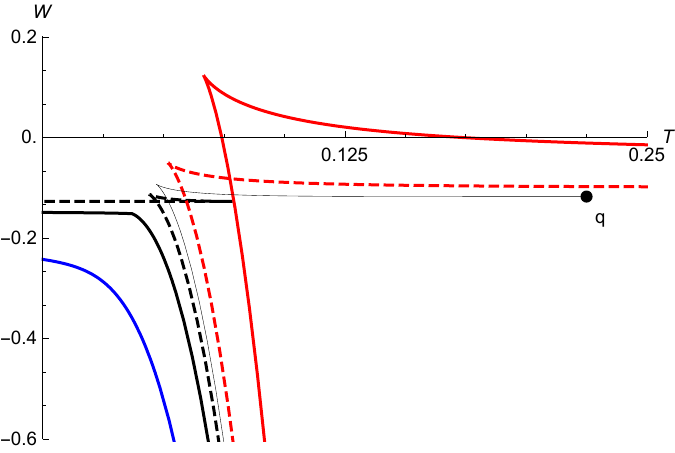}
	\caption{{\bf $W-T$ diagram: effect of $\phi$.} The grandcanonical free energy $W$ is plotted against the temperature $T$ for fixed $P=0.01$ and $\alpha=1$, and
	$\phi=0.5$ (red solid with $W_{\mbox{\tiny cusp}}>0$), 
$\phi=0.67$ (red dashed  with  $W_{\mbox{\tiny cusp}}<0$),	
$\phi=\phi_c=1/\sqrt{2}\approx 0.707$ (thin black curve with a termination point $q$), $\phi=0.725$ (black dashed with a swallowtail), $\phi=0.76$ (black solid with a critical point) and $\phi=0.85$ (smooth solid blue). As clearly seen from the figure, as $\phi$ varies, we observe qualitatively very different behavior that results in various types of phase transitions, as displayed in the following $\phi-T$  diagram.}
\label{fig15} 
\end{center}
\end{figure}

\begin{figure}
\begin{center}
%\begin{figure}[h]
	\includegraphics[scale=0.67]{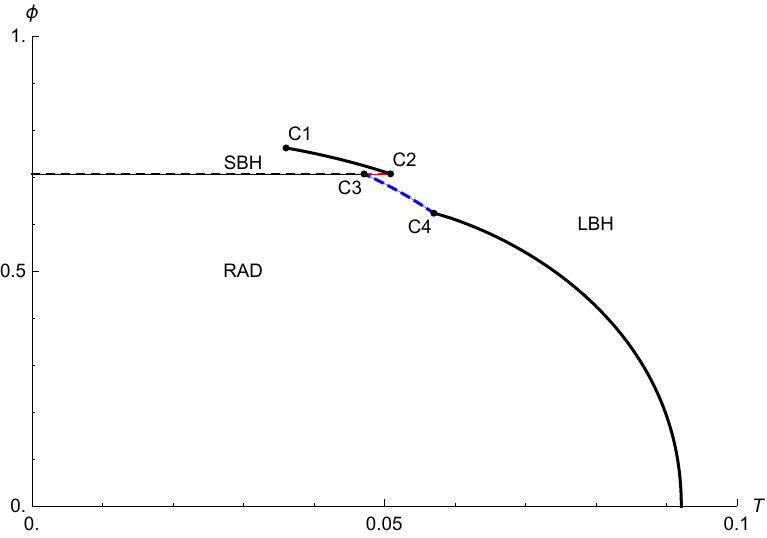}
	\caption{
\textbf{$\phi-T$ phase diagram.} The diagram is displayed for fixed $\alpha=1$ and $P=0.01$. We observe first order phase transitions between SBH and LBH phases  
(solid black curve from $C1$ to $C2$) and between the radiation (RAD) and LBH phases (solid black curve to the right of $C4$); here $C_1$ is a standard critical point where the phase transition becomes of the second-order. In addition we observe zeroth-order phase transitions between all phases (denoted by dashed red, black, and blue curves). The dashed red and black 0th-order transitions lie on the $\phi=\phi_c$ line -- they separate SBH/RAD and SBH/LBH phases, respectively. For small enough pressures, the 0th-order transition between radiation and large black hole phases (blue dashed curve between $C3$ and $C4$) would vanish and be replaced with the extended first-order phase transition going all the way to $\phi=\phi_c$. The diagram is symmetric w.r.t. the $\phi=0$ axis and the solid black coexistence curve is regular at $\phi=0$. 
	}
\label{fig16} 
\end{center}
\end{figure}

Let us finally turn to, perhaps even more interesting,  $\phi-T$ phase diagram,  where we let $\phi$ to vary for fixed $\alpha$ and $P$. The corresponding behavior of the $W-T$ diagram is for fixed $P=0.01$ and $\alpha=1$ displayed in Fig.~\ref{fig15}, while the associated $\phi-T$ phase diagram is displayed in Fig.~\ref{fig16}. We clearly see the importance of the critical $\phi_c$, which, when crossed, results in a zeroth-order phase transition for small enough temperatures between the radiation and small black hole phases and between small black holes and large black holes for intermediate temperatures. The phase diagram Fig.~\ref{fig16} is rather complex, and as far as we know unprecedented in black hole thermodynamics.  It clearly illustrates the complex thermodynamic behavior of the RegMax charged AdS black holes.\footnote{We remark that for small enough pressures, 
the behavior of $W-T$ slightly changes: we no longer have curves with $W_{\mbox{\tiny cusp}}<0$. Consequently, for such pressures,  the blue dashed 0th-order separation line between radiation and LBH in Fig.~\ref{fig16} is replaced by the extended first-order solid black curve, which now extends all the way from $\phi=0$ to $\phi=\phi_c$.}

\section{Other notable solutions}
\label{sec5}
In this section, we shall go beyond spherical symmetry, 
and present novel accelerating black holes 
in RegMax theory. We also review slowly rotating black holes, following \cite{Kubiznak:2022vft}, and 
construct  weakly NUT charged Taub-NUT solutions.
It is a remarkable property of the RegMax theory, that all these solutions can analytically be found and take a simple form that is in many ways very similar to what happens in the linear Maxwell case.   
We also refer the reader to \cite{Tahamtan:2020lvq} and \cite{Tahamtan:preparation} for the discussion of radiative Robinson--Trautman spacetimes in this theory, which are also remarkably Maxwell-like.

%%%%%%%%%%%%%%%%%%%%%%%%%%%%%%%%%%%%%%%
%%%%%%%%%%%%%%%%%%%%%%%%%%%%%%%%%%%%%%%
\subsection{Accelerated black holes}

In general relativity accelerated black holes are described by the so called {\em C-metric} and its generalizations \cite{Kinnersley:1970zw, Plebanski:1976gy, Dias:2002mi, Griffiths:2006tk, Podolsky:2022xxd, Cembranos:2022dhi}. 
This is a remarkable exact solution of the Einstein--Maxwell theory, an exact radiative spacetime, which can be used to describe 
rich physical phenomena, such as pair creation of black holes, e.g. \cite{Dowker:1993bt}, or can serve as a test playground for studying the properties of gravitational and electromagnetic radiation, e.g. \cite{Bicak:2002yk, Podolsky:2003gm}.
In the simplest case, 
the acceleration of the black hole is caused
by a conical deficit (or surplus) on one side of the hole.
In a more realistic setting, the conical singularity pulling the black hole is  replaced by a finite width cosmic string
core \cite{Gregory:1995hd}, or a magnetic flux tube \cite{Dowker:1993bt}, relating the 
acceleration to the interaction with a local cosmological medium.
In its turn, the acceleration produces gravitational and electromagnetic radiation which escapes to regions behind the acceleration horizon (if it exists). Because of the presence of conical deficits and acceleration horizons, the physical charges of the solution and its thermodynamics have only been understood quite recently, see \cite{Astorino:2016ybm, Appels:2016uha, Appels:2017xoe, Anabalon:2018ydc, Anabalon:2018qfv, Ball:2020vzo, Ball:2021xwt}, and also \cite{Cassani:2021dwa, Boido:2022iye}.

The standard (non-rotating) C-metric can be written in the following form, c.f. \cite{Appels:2016uha, Anabalon:2018qfv} (see also App.~\ref{appA} for an alternative coordinate system): 
\be\label{Cansatz}
ds^2=\frac{1}{\Omega^2}\Bigl(-fdt^2+\frac{dr^2}{f}+r^2\Bigl[\frac{dx^2}{h}+h \frac{d\varphi^2}{K^2}\Bigr]\Bigr)\,, 
\ee
where $f=f(r), h=h(x)$ are two metric functions of one variable only, $\Omega$ is the conformal factor, given by 
\be\label{OmegaHl}
\Omega=1-{\cal A}rx\,, 
\ee
${\cal A}$ is the acceleration parameter, and $K$ is a parameter controlling the conical deficit (while $\varphi$ is assumed to have periodicity of $2\pi$).

In the Einstein--Maxwell theory, the metric is accompanied by the following vector 
potential:
\be
A=-\frac{Q}{r}dt\,, 
\ee 
and the metric functions $f$ and $h$ take the following explicit form \cite{Appels:2016uha, Anabalon:2018qfv}:
\ba
f^M&=&(1-{\cal A}^2r^2)f_0^M+\frac{r^2}{\ell^2}\,,\quad \,,\\
h^M&=&(1-x^2)(1+2m{\cal A}x+{\cal A}^2Q^2x^2)\,,\label{hMmaxwell}
\ea
where $Q$ is the charge, $m$ is the mass parameter, and we have defined 
\be\label{f0M}
 f_0^M=1-\frac{2m}{r}+\frac{Q^2}{r^2}\,.
\ee
Obviously, the metric function $h^M$ is in a factorized form, and has two fixed roots at $x=\pm1$. Concentrating on the region in between them, 
%the  for positive $m$ and ${\cal A}$, function $h^M$ is positive for 
\be \label{xint}
x\in (-1,1)\,,
\ee
one can arrange that (for a certain proper choice of parameters) $h^M$ is positive, and 
the metric \eqref{Cansatz} describes  spherical accelerating black holes; in this region one can also set $x=\cos\theta$.

The charged C-metric has been generalized to various supergravities, e.g. \cite{Siahaan:2018qcw,
Cassani:2021dwa, Boido:2022iye,Ferrero:2021ovq, Wang:2022hzh, Nozawa:2022upa}.
However, a generalization to theories of NLE seems quite challenging, e.g., \cite{Breton:2023bwf}. In fact, the only known C-metric solution in NLE is for the ModMax theory \cite{Barrientos:2022bzm}, which, however, is in many aspects very similar to what happens in the Maxwell case. Here, we present highly non-trivial generalization of the standard charged C-metric in RegMax theory, see also App.~\ref{appA} for the presentation of this solution in a different coordinate system.

The charged AdS C-metric in the RegMax theory takes the form \eqref{Cansatz}, \eqref{OmegaHl},   where the metric functions now read
\ba
f&=&f_0-{\cal A}^2r^2f_0^M\,,\\
h&=&1+\frac{2x}{{\cal A}}(2\alpha^3\sqrt{|Q|}+{\cal A}^2m)
+({\cal A}^2Q^2+2\alpha^2|Q|-1)x^2\nonumber\\
&+&\frac{4\alpha |Q|^{3/2}-6m}{3}{\cal A}x^3 +\frac{4\alpha^4}{{\cal A}^2}\log\Bigl(1-\frac{{\cal A}x\sqrt{|Q|}}{\alpha}\Bigr)\,.\quad 
\ea
Here, $f_0$ is the static RegMax function \eqref{f-SSS}, $f_0^M$ is given by \eqref{f0M}, and the metric is accompanied by the following vector potential:
\be
A=-\frac{\alpha Q}{\alpha r+\sqrt{|Q|}}dt\,,
\ee
which is formally identical to \eqref{phi}.
With these, it can easily be shown that the above Maxwell solution is recovered upon $\alpha\to \infty$, while ${\cal A}\to 0$ limit yields the spherical solution discussed in the previous section.

\begin{figure}
\begin{center}
\includegraphics[scale=0.67]
{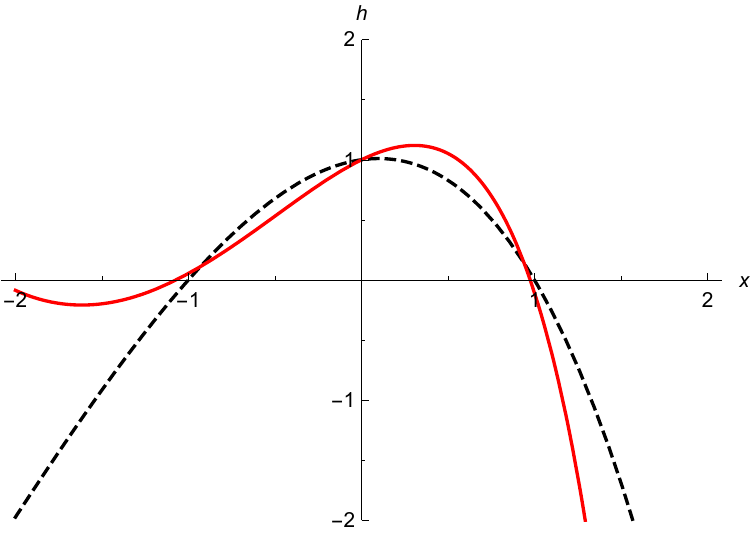}
	\caption{{\bf Metric function $h$.} We display the metric function $h$ as a function of $x$ for $\alpha=1, m=1, Q=1.35, \ell=4$ and i) slowly accelerating black hole with ${\cal A}=0.1$ denoted by dashed black curve and ii) fast accelerating black hole with ${\cal A}=0.35$ denoted by red solid line. Clearly the two roots $x_\mp$, outlining the region in which $h$ is positive, slightly vary with the choice of parameters. It is this fact that complicates the calculation of thermodynamic charges of this solution.    
}
\label{fig17} 
\end{center}
\end{figure}

\begin{figure}
\begin{center}
\includegraphics[scale=0.67]
{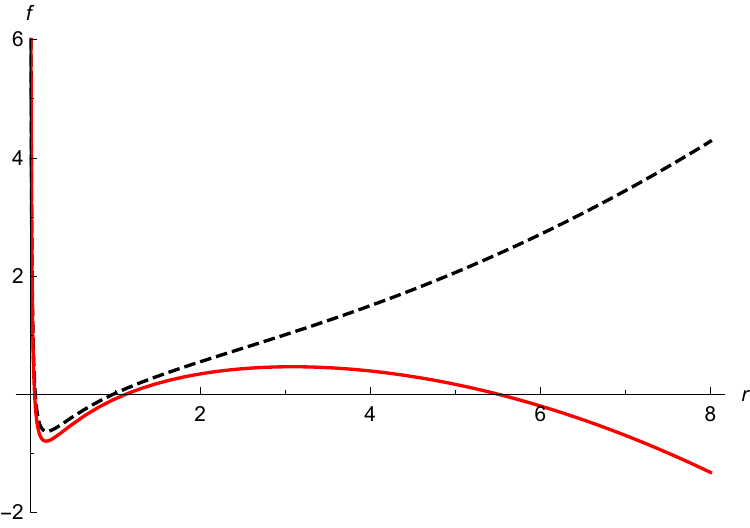}
	\caption{{\bf Horizons of RegMax C-metric.}
We display the metric function $f$ as a function of $r$ for slowly accelerating black holes (dashed black) and fast accelerating black holes (solid red), for the same choice of parameters as in  Fig.~\ref{fig17}. Obviously, the fast accelerating case features the presence of an additional (acceleration) horizon.}
\label{fig18} 
\end{center}
\end{figure}

In order for our new metric to describe spherical accelerated black holes in AdS,   we need to restrict to a region 
between two finite roots of $h$, where $h$ is positive. Contrary to the Maxwell case, this is no longer given by \eqref{xint}, and is, due to the presence of the logarithmic term, hard to determine analytically.  In what follows, we assume that such region exists, denoting the corresponding roots $x_\mp$, respectively:
\be
x\in(x_-, x_+)\,. 
\ee
These are no longer constant, but rather depend on the chosen parameters characterizing the solution, as displayed for $\alpha=1$ (deep RegMax regime) in Fig.~\ref{fig17}. Since thermodynamic charges depend on $x_\mp$, this fact complicates the study of thermodynamics of these black holes. Another complication stems from the existence of various horizons, as determined from the roots of metric function $f$, which now also contains a logarithm. It is known that in the Maxwell AdS case, there exists a {\em slowly accelerating regime}, where no acceleration horizon nor radiation are present. For such black holes thermodynamics can easily be formulated, see 
\cite{Appels:2016uha, Appels:2017xoe, Anabalon:2018ydc, Anabalon:2018qfv}. On the other hand, when the acceleration horizon, and possibly other horizons as well, are present, in the so called {\em fast accelerating regime}, the thermodynamics is not so straightforward, see however \cite{Ball:2020vzo, Ball:2021xwt}. In Fig.~\ref{fig18}, we display the behavior of $f$ for $\alpha=1$ and two choices of the acceleration parameter. 
This figure clearly illustrates that 
both (slowly accelerating and fast accelerating) regimes remain present even in the deep RegMax regime. We plan to return to the discussion of the admissible parameter space of this newly found solution and the discussion of its thermodynamics in future work \cite{TDsprep}.

To conclude this section, let us stress that 
it is a remarkable feature of the RegMax theory, that the charged AdS C-metric can be written in the standard form \eqref{Cansatz}, with the conformal pre-factor $\Omega$ given by \eqref{OmegaHl}. While we were not able to check this explicitly, it is reasonable to conjecture that apart from the ModMax and Maxwell theories, RegMax is  
the only restricted NLE for which the C-metric can be written in this form.

\subsection{Slowly rotating AdS black holes}
Constructing a NLE generalization of the Kerr--Newman solution remains a holy grail of NLE studies. So far, no such (fully analytic) solutions are known. This is partly because, as shown in \cite{Kubiznak:2022vft}, one cannot use the Newmann--Janis trick to generate them. Such a trick works in the Maxwell case and was used to generate the Kerr--Newmann solution starting from the Reissner--Nordstr\"{o}m one \cite{Newman:1965tw}. However, as shown in the above paper, in the case of NLE the Newmann--Janis trick already fails at the linear order in rotation parameter $a$. For this reason, it is instructive to at least construct slowly rotating black holes in NLE \cite{Kubiznak:2022vft}.  
For convenience, we briefly repeat here the corresponding discussion.

Namely, starting from a spherical solution in any NLE, one can find a slowly rotating one (valid to the linear order in rotation parameter $a$) by employing the following ansatz for the 
metric and for the vector potential: 
\ba
ds^2&=&-f_0dt^2+\frac{dr^2}{f_0}+2ar^2\sin^2\!\theta h dt d\varphi +r^2 d\Omega^2\,,\quad\\
A&=&\psi_0 (dt-a\omega \sin^2\!\theta d\varphi)\,, 
\ea 
where $f_0$ and $\psi_0$ are the spherical metric function and the electrostatic potential, respectively, and $\omega=\omega(r), h=h(r)$ are two new functions encoding the effect of rotation.
Unfortunately, for a generic NLE, the two new functions obey complicated differential equations, see \cite{Kubiznak:2022vft}, and cannot be solved for explicitly.

In this respect, RegMax has a privileged position. Namely, as shown in \cite{Kubiznak:2022vft}, RegMax is the only NLE apart from Maxwell, that is fully characterized by invariant ${\cal S}$ and whose slowly rotating solutions can be written in the above form with 
\be
\omega=1\,. 
\ee  
In other words, RegMax and Maxwell are the only restricted NLE's whose slowly rotating vector potential is fully characterized by the electrostatic potential $\psi_0$. Interestingly, the proof of this statement 
is constructive -- this is how the Lagrangian \eqref{Tay} was derived in \cite{Kubiznak:2022vft}.

To write the slowly rotating solution explicitly, we employ the functions $f_0$ and $\psi_0$ given by \eqref{f-SSS} and \eqref{phi}.
Moreover, the function $h$ can be explicitly found \cite{Kubiznak:2022vft} and reads 
\be
h=\frac{f_0-1}{r^2}-\frac{2\sqrt{|Q|}}{3\alpha r^3}\,, 
\ee
completing thus the solution.

It would be very interesting to see, whether this approximate solution can be extended to a fully rotating black hole in the RegMax theory -- giving thus a first example of a rotating solution in theories of NLE.

\subsection{Taub-NUT solution}
In principle one can also find a Taub-NUT solution, using the procedure outlined in \cite{Kubiznak:2022vft}. It takes the following standard form:
\ba
ds^2&=&-f(dt+2n\cos\theta d\varphi)^2+\frac{dr^2}{f}+(r^2+n^2)d\Omega^2\,,
%\quad\label{TNmetric}
\nonumber\\
A&=&\psi(dt+2n\cos\theta d\varphi)\,,\label{TNvector}
\ea
where $n$ denotes the NUT parameter. 
The RegMax equations together with the Einstein equations then yield ordinary differential equations for $f(r)$ and $\psi(r)$. Unfortunately, it turns out that for the RegMax theory these are difficult to solve and we were unable to find the full analytic solution.

For this reason, let us only present the `weakly NUT charged' Taub-NUT spacetime, a solution valid to linear order in $n$. This is simply given by the above ansatz \eqref{TNvector} (with $n^2$ neglected in the metric), upon using the static metric function $f=f_0$ \eqref{f-SSS}, together with the static potential $\psi=\psi_0$, \eqref{phi}. One can easily check that this solves all equations to at least $O(n)$ order.

%%%%%%%%%%%%%%%%%%%%%%%%%%%%%%%%%%%%%%%%%%%%%
%%%%%%%%%%%%%%%%%%%%%%%%%%%%%%%%%%%%%%%%%%%%%%
\section{Magnetically charged solutions}
\label{sec6}
In order to construct magnetically charged solutions in the RegMax theory we need to generalize the Lagrangian \eqref{Tay}, to accommodate for the possibility of ${\cal S}>0$ while keeping $\alpha$ positive. To this purpose, we start with the 
Lagrangian ${\cal L}(s({\cal S}))$, \eqref{Tay} and \eqref{s}, which is valid only for ${\cal S}<0$, and define new  Lagrangian by:
\be\label{LMag}
\tilde {\cal L}({\cal S})=-\sgn({\cal S})\, {\cal L}\big(-\sgn({\cal S})\, s(-\sgn({\cal S})\,{\cal S})\big)\,.
\ee
This has the following derivatives:
\be
\tilde {\cal L}_{{\cal S}}={\cal L}_{{\cal S}}\,,\quad \tilde {\cal L}_{{\cal SS}}=-\sgn({\cal S})\, {\cal L}_{{\cal SS}}\,,
\ee
which are well-defined when we assume the regularity of derivatives of ${\cal L}$ (we have used ${\cal L}|_{{\cal S}=0}=0$ and ${s}|_{{\cal S}=0}=0$ in the derivation above). Furthermore, we see that the sign of the first derivative does not flip when ${\cal S}$ changes sign but for the second derivative it seemingly does. However, considering the dependence on invariant ${\cal S}$ (as seen in \eqref{LMag}) we observe that $\tilde{\cal L}_{{\cal S}}<0$ and  $\tilde{\cal L}_{{\cal SS}}>0$ irrespective of the sign of ${\cal S}$. This means that the discussion of birefringence in section \ref{sec3} (see mainly Eq.~\eqref{null-norm} and the text thereafter) is valid for $\tilde{\cal L}$ and the propagation of modes is causal as desired.

\subsection{Spherical solutions}

To simplify the matters discussed above, we can take the Lagrangian \eqref{Tay}, redefine $s$, and flip the overall sign, to obtain 
\ba\label{TayMag}
\tilde {\cal L}&=&2\alpha^4 \Bigl(1-3\ln(1-s)+\frac{s^3+3s^2-4s-2}{2(1-s)}\Bigr)\,,\nonumber\\
s&=&-\Bigl(\frac{\mathcal{S}}{\alpha^4}\Bigr)^\frac{1}{4}\,,
\ea 
and take again $\alpha>0$. 
This, as we shall see, leads to a well-defined solution with correct Maxwell limit \footnote{Interestingly, in the strong field regime the magnetic RegMax Lagrangian behaves like $\sqrt{{\cal S}}$, similar to the Born--Infeld case (for magnetic solutions). Such `square root Lagrangian' was already studied 50 years ago \cite{Nielsen1973} and more recently in \cite{Tahamtan-Kundt:2017,Tahamtan-PRD:2020}.}. Namely, we assume the following 
local potential 
\be
A=Q_m\cos\theta d\varphi\,, 
\ee
with the corresponding field strength
\be 
F= Q_m \sin{\theta} d\varphi \wedge d\theta\,,
\ee
for which ${\cal S}=\frac{2Q_m^2}{r^4}$ and ${\cal P}=0$.
The RegMax equations, \eqref{FE}, are then automatically satisfied, and from the Einstein equations \eqref{Hmunu}, the metric ansatz \eqref{SSS} yields 
\ba\label{Magnetic-NEW-SSS}
f_0&=&1-2\alpha^2|Q_m|+\frac {4\alpha|Q_m|^{3/2}-6M}{3r}+\frac{r^2}{\ell^2}\nonumber\\
&&+4\sqrt{|Q_m|}\alpha^{3}r-4{\alpha}^{4}{r}^{2}
	\ln  \Bigl(1+\frac{\sqrt{|Q_m|}}{r\alpha}\Bigr)\,, 	
\ea
where 
\be
Q_m=-\frac{1}{4\pi}\int_{S^2} F\, 
\ee 
is the (asymptotic) magnetic charge, and $m$ is the mass parameter (see below).
This has the following large $\alpha$ expansion:
\be
f_0=1-{\frac {2m}{r}}+\frac{Q_m^2}{r^{2}}+\frac{r^2}{\ell^2}+O\left(\frac{1}{\alpha}\right)\,,
\ee
and we recovered the magnetically charged AdS black hole solution in Maxwell's theory in the appropriate limit.

The constructed solution is characterized by the following thermodynamic variables:
\ba
M&=&m\,,\quad T=\frac{f'(r_+)}{4\pi}\,,\quad S=\pi r_+^2\,,\nonumber\\
\phi_m&=&\frac{\alpha Q_m}{\alpha r_++\sqrt{|Q_m|}}\,, \quad 
P=\frac{3}{8\pi \ell^2}\,,\quad  V=\frac{4}{3}\pi r_+^3\,,\nonumber\\
\mu_\alpha&=&
\frac{2}{3}\frac{Q_m^2-2|Q_m|^{3/2}\alpha r_++12\alpha^3r_+^3\sqrt{|Q_m|}+6|Q_m|\alpha^2r_+^2}{r_+\alpha+\sqrt{|Q_m|}}\nonumber\\
&&-8\alpha^3 r_+^3\log\Bigl(1+\frac{\sqrt{|Q_m|}}{r_+\alpha}\Bigr)\,,
\ea
which satisfy the corresponding extended first law and the generalized Smarr relation:
\ba
\delta M&=&T\delta S+\phi_m \delta Q_m+V\delta P+\mu_\alpha \delta \alpha\, \\
M&=&2TS+\phi_m Q_m-2VP-\frac{1}{2}\mu_\alpha \alpha\,. 
\ea
Here, the magnetic potential was obtained with the help of the dual vector potential $\tilde A$, defined by $d\tilde A=*D$. Namely, 
\be
\phi_m=-\xi\cdot \tilde A\Bigl|_{r=r_+}\,, 
\ee
where 
\be
\tilde A=-\frac{\alpha Q_m}{r\alpha+\sqrt{|Q_m|}}dt\,. 
\ee
Obviously, at spherical level, the magnetic solutions are characterized by the same metric function and possess the same thermodynamic quantities as the electric ones, upon replacing $Q\leftrightarrow Q_m$. As we shall see now, however, this symmetry is broken beyond the spherical case.

\subsection{Slowly rotating solutions}

As shown in \cite{Kubiznak:2022vft}, spherical magnetic solutions in NLE may be upgraded to slowly rotating ones  by employing the following ansatz:
\ba
ds^2&=&-f_0 dt^2+\frac{dr^2}{f_0}+2ar^2\sin^2\!\theta h dt d\varphi +r^2 d\Omega^2\,,\\
A&=&Q_m\cos\theta\Bigl(d\varphi-\frac{a\omega}{r^2}dt\Bigr)\,, 
\ea
where $\omega(r)$ and $h(r)$ are new vector potential and metric functions, respectively. Here we construct the slowly rotating solution in the RegMax theory explicitly.

Namely, it can be shown that a consistent solution can be found in the above form, where we set 
\be
\omega=1+\frac{2\sqrt{|Q_m|}}{3r\alpha}\,, 
\ee
$f_0$ is given by \eqref{Magnetic-NEW-SSS}, and
\be
h=\frac{f_0-1}{r^2}-\frac{2\sqrt{|Q_m|}}{3r^3\alpha}\,. 
\ee
As we see, and contrary to the slowly rotating electric case discussed in the previous section, the vector potential now picked up a correction at the order $1/\alpha$, and thence it is not simply  given only in terms of the magnetostatic potential. To prevent a confusion, we stress that the above is a slowly rotating solution, which, however, is valid to any order in $1/\alpha$.

\subsection{Magnetic Taub--NUT}
In order to construct a magnetically charged Taub-NUT solution to linear order in the NUT parameter $n$, let us first discuss the form of the ansatz, as inspired by what happens in the Maxwell theory. For this, let us start from the full magnetic Taub-NUT in linear Maxwell electrodynamics, e.g. \cite{Bordo:2019slw}, which takes the form \eqref{TNvector}, where 
\ba
f&=&\frac{r^2-2mr-n^2+4n^2g^2}{r^2+n^2}-\frac{3n^4-6n^2r^2-r^4}{\ell^2(r^2+n^2)}\,,\quad\\
\psi&=&-g\frac{r^2-n^2}{r^2+n^2}\,,
\ea  
and the magnetic charge is given by $Q_m=-2ng$.

The limit of the vanishing NUT charge is obtained by setting $n\to 0$ and $g\to \infty$ so that the magnetic charge $Q_m$ remains finite. For the metric function $f$ this yields the static solution. However, in order for this limit to make sense also for the vector potential, one has to add a gauge term, $A\to A+gdt$, which yields  
\be
A=-\frac{nQ_m}{r^2+n^2}dt+Q_m\frac{r^2-n^2}{r^2+n^2}\cos \theta d\varphi\,, 
\ee 
and,  to the linear order in $n$, gives 
\be\label{vectorTNsmall}
A=-nQ_m \nu dt+Q_m\cos\theta d\varphi\,, 
\ee
where 
\be \label{psiNUTmag}
\nu=\frac{1}{r^2}
\ee
for the Maxwell case.

Following the above discussion, we seek the magnetized weakly NUT charged solution in the form \eqref{TNvector} for the metric, together with the vector potential ansatz \eqref{vectorTNsmall}. For the RegMax theory this yields $f=f_0$, \eqref{Magnetic-NEW-SSS}, together with:
\be 
\nu=
\frac{1}{r^2}+\frac{2\sqrt{|Q_m|}}{3\alpha r^3}\,,
%\frac{3r\alpha-2\sqrt{|Q_m|}}{3\alpha r^3}\,,
\ee
which has the right Maxwell limit in the large $\alpha$ expansion.

\section{Summary}
\label{sec7}
 In this paper we have studied 
the basic properties and solutions of a particular model of non-linear electrodynamics, which we named the RegMax theory. Such a theory is characterized by a dimensionfull parameter $\alpha$ and is neither conformal nor it possesses an electromagnetic duality. However, while the 
RegMax Lagrangian \eqref{Tay} seems rather complicated at first sight, it
leads to arguably the most straightforward regularization of the electric field of a point charge \eqref{E}. Even more importantly, the RegMax theory becomes truly 
remarkable when the self-gravitating solutions are considered. 
Namely, apart from the radiative spacetimes of the Robinson--Trautmann class \cite{Tahamtan:2020lvq} and the slowly rotating solutions 
\cite{Kubiznak:2022vft}, the RegMax model provides further important gravitating solutions.

Perhaps the biggest discovery  regarding the exact solutions in this theory so far is the hereby presented C-metric (although its existence could have been anticipated from the results of \cite{Tahamtan:2020lvq}).  Similar to other exact spacetimes in this theory, the overall structure of this solution is remarkably Maxwell-like and clearly generalizes the standard charged C-metric of the Einstein--Maxwell theory. 
It is tempting to conjecture that the RegMax theory is the only restricted NLE for which the solution for accelerated AdS black holes
can be found in this form.

%{
%Namely the exact C-metric solution representing accelerated AdS black holes is rather rare in Einstein--NLE system. Its overall structure clearly generalizes that of the Maxwell C-metric geometry. Complete analysis of this solution and its thermodynamic properties will be subject of further investigation.

The original Lagrangian \eqref{Tay} is not directly applicable to purely magnetic solutions. For this reason, 
we have `naturally' extended the RegMax theory in 
Sec.~\ref{sec6} to arrive at \eqref{LMag}, and in particular at \eqref{TayMag}, allowing us to derive the magnetically charged black holes and their slowly rotating cousins. However, contrary to the slowly rotating electric solutions, the electromagnetic field of the latter is  no longer entirely governed by the static potential, but rather picks up a $1/\alpha$ correction, departing thus from the `Maxwellness' of the theory. It remains to be seen whether another extension, more aligned with the observed `Maxwell spirit' of the electric theory, can be formulated.

We have also devoted a large body of our  work to analyzing the thermodynamic behavior of electrically charged spherically symmetric AdS black hole solutions. In the canonical (fixed charge) ensemble, this leads to the standard Van der Waals like behavior for large $\alpha$ (Maxwell-like regime) and to the Schwarzschild-like behavior for small $\alpha$. Interestingly, and contrary to what happens in the Born--Infeld case \cite{Gunasekaran:2012dq}, for RegMax there is no intermediate range of $\alpha$'s for which one would observe multicomponent behavior of reentrant phase transitions. It remains to be seen, which behavior, whether the one with the intermediate region (as in the Born--Infeld case) or the one without it (RegMax theory) is more generic in theories of NLE.

Perhaps even more interesting is the thermodynamic behavior in the grandcanonical (fixed potential) ensemble, for which the main results can be read off from a rather  complex phase diagram in Fig.~\ref{fig16}. This figure illustrates the presence of multiple first-order, second-order, and zeroth-order phase transitions between radiation, small black hole, and large black hole phases.  
%Thermodynamics of magnetically charged black holes is also interesting and will be discussed elsewhere \cite{magPrep}.

To summarize, we have shown that the RegMax theory is a rather interesting NLE, especially at the level of self-gravitating solutions. The newly obtained exact solutions  will provide a playground for studying strong electromagnetic fields in non-trivial (beyond spherical symmetry) gravitating backgrounds. The key question remaining is, whether the RegMax theory remains remarkable only at the level of its gravitating solutions, or, whether it features some more fundamental properties as well. 
For example, can the Maxwellness of the electric solutions be also carried over to the magnetic ones? Is it possible to extend the Lagrangian \eqref{Tay} to include the invariant ${\cal P}$, so that the electromagnetic duality would be restored? 
Or perhaps even more interestingly, can the (possibly generalized) RegMax theory, similar to the Born--Infeld case, be derived from some more fundamental (possibly higher-dimensional) theory?

\subsection*{Acknowledgements}
We would like to thank Pavel Krtou{\v s} for numerous discussions. D.K. 
is grateful for support from GA{\v C}R 23-07457S grant of the Czech Science Foundation. T.T. was supported by Research Grant No. GA\v{C}R 21-11268S and O.S. by Research Grant No. GA\v{C}R 22-14791S.

\appendix

\section{C-metric in $x-y$ coordinates}
\label{appA}
The C-metric is most easily written in the standard ``$x-y$ coordinate system". In this appendix, we present the corresponding ansatz for the Maxwell theory and show that it can also be applied in the RegMax case.

The standard charged C-metric ansatz in $x-y$ coordinates is given by
\be\label{XYansatz}
ds^2=\frac{1}{H^2\!(x,y)}\Bigl(-F(y)dt^2+\frac{dy^2}{F(y)}+\frac{dx^2}{G(x)}+G(x)d\varphi^2\Bigr)\,, 
\ee
where $G=G(x)$ and $F=F(y)$ are two metric functions and $H=H(x,y)$ is a conformal factor. With this ansatz, the off-diagonal Einstein tensor reads
\begin{equation}\label{offDiagonal-Cmetric}
G_{xy}=2 \frac{H_{,xy}}{H}\,.	
\end{equation}

\subsection{Electrically charged case}

Let us first focus on the electrically charged case, accompanying the metric with the following vector potential:
\be \label{electric charged-Cmetric}
A=\psi(y)dt\,. 
\ee
It follows that the invariant ${\cal S}$ reads 
\be\label{CmetricS}
 {\cal S}=-H^4 (\psi_{,y})^2\,.
\ee
By integrating once the modified Maxwell equation, $(\nabla~\cdot~D)_t=0$, we obtain:
\begin{equation}\label{psiprime}
		\psi_{,y}=\frac{c(x)}{\mathcal{L}_{S}(y,x)}\,,
	\end{equation}
	where $c(x)$ is an integration `constant'. In case of the Maxwell theory, the term $\cal{L}_S$ is constant, and  therefore $c(x)$ must be constant as well. For any other theory $\cal{L}_S$ should be (multiplicatively) separable in $y$ and $x$. Together with the fact that $\cal{L}_S$ is a function of ${\cal S}$, which takes the above form \eqref{CmetricS} (see also \eqref{HxHy} below), Eq.~\eqref{psiprime} imposes a very non-trivial restriction. Surprisingly, as we shall see below, this condition is satisfied for the RegMax theory.

Before specifying to a concrete NLE model, let us conclude with an important observation. It follows from the form of the  vector potential \eqref{electric charged-Cmetric}, that all the off diagonal components of the energy-momentum tensor have to vanish. In particular, 
\be 
T_{xy}=0\,, 
\ee
which together with \eqref{offDiagonal-Cmetric} implies that $H_{,xy}=0\,.$ In other words, $H$ has to additively separate: 
\be\label{HxHy}
H=H_x(x)+H_y(y)\,.
\ee
To see what are the consequences of various choices for conformal factors, let us first look at what happens in the Maxwell theory.

\subsection{Maxwell electrodynamics}
In the Maxwell case we find the following three interesting solutions, depending on the choice of the conformal factor $H\in\{1, y, x+y\}$. The first solution:
\ba
H&=&1\,,\nonumber\\ 
F&=&Q^2y^2+\frac{3y^2}{\ell^2}+c_1y+c_2\,,\\ 
G&=&-Q^2x^2+\frac{3x^2}{\ell^2}+c_3x+c_4\,,\nonumber
\ea
%\ba
%H&=&1\,,\nonumber\\ 
%F&=&2Q^2y^2+c_1y+c_2\,,\\
%G&=&-2Q^2x^2+c_3x+c_4\,,\onumber
%\ea
is accompanied by the following vector potential 
\be\label{psixy}
A=\psi dt\,,\quad \psi=-Qy\,. 
\ee
It is easy to show that in this case the standard curvature invariants, as well as ${\cal S}$ are constant, and the spacetime describes a maximally symmetric space with `uniform electric field' that mimics the cosmological constant.   

The second choice,  
\ba
 H&=&y\,, \nonumber\\
 F&=&y^2\Bigl(Q^2y^2+c_1y+c_2+\frac{1}{\ell^2 y^2}\Bigr)\,,\\
 G&=&-c_2 x^2+c_3 x+c_4\,, \nonumber
\ea
accompanied by the `same $A$', \eqref{psixy}, yields, upon the due change of coordinates and choice of parameters the standard spherical charged AdS black hole solution. 

Finally, for $A$ given by \eqref{psixy} and choosing 
\ba 
H&=&x+y\,,\nonumber\\
F&=&Q^2y^4+\frac{1}{6}c_1y^3-\frac{1}{2}c_2y^2+c_3y-c_4+\frac{1}{\ell^2}\,,\qquad\\
G&=&-Q^2x^4+
\frac{1}{6}c_1x^3+\frac{1}{2}c_2x^2+c_3x+c_4\,,\nonumber
\ea
we obtain the charged C-metric. In particular, 
upon changing coordinates according to 
\be\label{trApp}
y=\frac{1}{r}\,,\quad x\to {\cal A} x\,,\quad \varphi\to \frac{\varphi}{K}\,, 
\ee
while introducing new metric functions: 
\be\label{trApp2} 
\Omega=r H\,,\quad f=r^2F\,,\quad h=\frac{G}{{\cal A}^2}\,,
\ee 
and for the following choice of the parameters:
\ba 
c_1&=&-12m\,,\quad c_2=2({\cal A}^2Q^2-1)\,,\nonumber\\
c_3&=&2{\cal A}^2 m\,,\quad c_4={\cal A}^2\,.
\ea
we recover the Maxwell AdS C-metric \eqref{Cansatz}--\eqref{hMmaxwell} presented in the main text, see also \cite{Griffiths:2009dfa} for more details on the C-metric.

\subsection{RegMax theory}
The calculations in the RegMax electrodynamics proceed in the same way. In particular, we also find three interesting solutions upon choosing $H\in\{1,y, x+y\}$. 

First, for   
\be
H=1\,, \quad A=-Qydt\,,
\ee
we find quadratic $F$ and $G$, and recover 
the maximally symmetric space. 

Second, when 
\ba
H&=&y\,,  \nonumber\\
F&=&4\sqrt{|Q|}\alpha^3\,y-4\alpha^{4}\log\Bigl(1+\frac{\sqrt{|Q|}y}{\alpha}\Bigr)\nonumber\\
&&+c_1 y^3+c_2y^2+\frac{1}{\ell^2}\,,\nonumber\\
G&=&-(2Q\alpha^2+c_2)x^2+c_3x+c_4\,, 
\ea
and for 
\be\label{Aapp}
 A=-\frac{\alpha\,Q\,y}{\alpha+\sqrt{|Q|}y}dt\,,
\ee
we have the spherically symmetric solution. Namely, by setting $y=1/r, \,x=\cos\theta$, $f_0=r^2F$, and choosing 
\be
c_{1}=\frac{4\alpha|Q|^{3/2}-6m}{3},\ c_{2}=1-2\alpha^{2}|Q|,\ c_{3}=0,\ c_{4}=1\,,
\ee
we recover the standard form of the spherical solution presented in the main text, \eqref{f-SSS}.

Third, setting 
\ba
H&=&x+y\,,\nonumber\\
F&=&-4\alpha^4\log\Bigl(1+\frac{\sqrt{|Q|}y}{\alpha}\Bigr)+\frac{c_1}{6}y^3-\frac{c_2}{2}y^2\\
&&+c_3y-c_4+\frac{1}{\ell^2}\,,\nonumber\\
G&=&4\alpha^4\log\Bigl(1-\frac{\sqrt{|Q|}x}{\alpha}\Bigr)+\frac{c_1}{6}x^3+\frac{c_2}{2}x^2+c_3x+c_4\,,\quad\nonumber
\ea
together with \eqref{Aapp}, we recover the RegMax AdS C-metric. Note that when the cosmological constant vanishes, the functions $F(y)$ and $G(x)$ have the following property: $F(w)=-G(-w)$. As discussed in the main text,  to maintain a Lorentzian signature of the metric \eqref{XYansatz}, it is necessary
	that $G > 0$, which implies that the coordinate $x$ must be constrained to lie	between appropriate (finite) roots of  function $G$.
	
To recover the form of the C-metric presented in the main text, we use \eqref{trApp} together with \eqref{trApp2}, and make the following choice of the integration constants:
\ba
c_1&=&8\alpha|Q|^{3/2}-12m\,,\quad
c_2=2({\cal A}^2Q^2+2|Q|\alpha^2-1)\,,\nonumber\\  
c_3&=&4\alpha^3\sqrt{|Q|}+2{\cal A}^2m\,,\quad c_4={\cal A}^2\,.
\ea 	

Finally, as we mentioned above, Eq.~\ref{psiprime} imposes a very strong restriction on NLE Lagragians, which seems almost impossible to satisfy. Surprisingly, for the RegMax theory and its vector potential \eqref{Aapp}, we find 
\begin{equation}
	{\cal L}_S=-\frac{(\alpha+y\sqrt{|Q|})^{2}}{(\alpha-x\sqrt{|Q|})^{2}}\,,
\end{equation}
which indeed is in the multiplicative separated form. Although we were not able to prove this rigorously,  we conjecture that RegMax is the only special NLE for which this happens.

\subsection{What about magnetically charged C-metric?}
Let us finally briefly comment on the purely magnetic C-metric. For the Maxwell theory, this is easily found in the form \eqref{XYansatz}, where the conformal factor 
reads
\be\label{Hxplusy}
H=x+y\,, 
\ee
$F$ and $G$ are quartic polynomials, and
the metric is accompanied by the following vector potential:
\be\label{AmagApp}
A=\psi(y) x d\varphi\,, 
\ee 
where $\psi$ has to be a constant, given by the magnetic charge, 
\be\label{psiQm} 
\psi=Q_m\,.
\ee 

Using the same ansatz \eqref{XYansatz}, \eqref{Hxplusy}, and \eqref{AmagApp} for any NLE beyond Maxwell, Eq.~\eqref{offDiagonal-Cmetric} yields $G_{xy}=0$, 
and therefore $T_{xy}$ has to be zero as well, i.e.,
\be 
T_{xy}=-4\frac{x\,(x+y)^2\,\psi\,\mathcal{L}_{S}}{G}\psi_{,y}=0\,,
\ee
that is, \eqref{psiQm} has to remain true. 
However, the modified Maxwell equation then requires 
\be
Q_m  \frac{\partial \mathcal{L}_{S}}{\partial x}=Q_m \mathcal{L}_{SS}\,{\cal S}_{,x}=0\,,
\ee
and since ${\cal S}=Q_m^2H^4$, this is a contradiction with the assumption. In other words, we just proved that, if purely magnetic C-metric is to exist in any NLE, it has to take a form which goes beyond our ansatz \eqref{XYansatz}, \eqref{Hxplusy}, and \eqref{AmagApp}.    
%Note that his would indicate that regular black hole solutions built using magnetic monopoles cannot be generalized towards C-metric geometry.}

\newpage

%\bibliography{references,myRef}
%\bibliography{references}
%\bibliographystyle{JHEP}

\providecommand{\href}[2]{#2}\begingroup\raggedright\endgroup

\end{document}